\documentclass[structabstract]{aa}

\usepackage{txfonts,epsfig,graphicx,natbib,url,twoopt}
\usepackage[breaklinks=true]{hyperref} %% to avoid \citeads line fills
\bibpunct{(}{)}{;}{a}{}{,}    %% natbib format like A&A and ApJ
\newcommandtwoopt{\citeads}[3][][]{\href{http://adsabs.harvard.edu/abs/#3}%
                                        {\citealp[#1][#2]{#3}}}
\newcommandtwoopt{\citepads}[3][][]{\href{http://adsabs.harvard.edu/abs/#3}%
                                        {\citep[#1][#2]{#3}}}
\newcommandtwoopt{\citetads}[3][][]{\href{http://adsabs.harvard.edu/abs/#3}%
                                        {\citet[#1][#2]{#3}}}
\newcommandtwoopt{\citeyearads}[3][][]%
   {\href{http://adsabs.harvard.edu/abs/#3}{\citeyear[#1][#2]{#3}}}

\begin{document}

%% ---------------------------------------------------------------------------
%% ---------------------------------------------------------------------------
%% The manuscript header -----------------------------------------------------

%% ------
%% Title
%% ------
%% <example>
%%
%%  \title{Optimality relationships for $p$-cyclic SOR p
%%  \thanks{Research supported in part by the US Air Force 
%% 		under grant no. AFOSR-88-0285 and 
%%		the National Science Foundation under grant 
%%		no. DMS-85-21154}\fnmsep 
%%  \thanks{This is a second footnote}\\ 
%%		resulting in asymptotically faster convergence\\ 
%%		for the same amount of work per iteration} 
%%  \subtitle{II. An example text with infinitesimal
%%		scientific value\\ 
%%  	whose title and subtitle may also be split} 

\title{Near-infrared properties of asymptotic giant branch stars 
	in nearby dwarf elliptical Galaxy NGC 205
	\thanks{Based on observations carried out at the Canada-France-Hawaii Telescope, 
 		operated by the National Research Council of Canada, 
		the Centre National de la Recherche Scientifique de France, 
		and the University of Hawaii.}}
%% -----------------
%% Names of authors
%% -----------------
%% <example>
%%
%%	\author{Daniel J. Pierce\inst{1}
%%	\and Apostolos Hadjidimios\inst{2} 
%%  \thanks{\emph{Present address:} 
%%    Department of Computer Science, Purdue University, 
%%    West Lafayette, IN 47907, USA} 
%%     \and Robert J. Plemmons\inst{3}} 
%% \offprints{R. Plemmons, \email{plemmons@...}}
%% \institute{Boeing Computer Service, P.O. Box 24346,
%%  MS 7L-21, Seattle, WA 98124-0346, USA 
%%  \and Department of Mathematics, University of Ioannina, 
%%  GR-45 1210, Ioannina, Greece 
%%  \and Department of Computer Science and Mathematics, 
%% North Carolina State University, Raleigh, NC 27695-8205, USA} 

\author{M.Y. Jung\inst{1} \and Jongwan Ko \inst{2,3} 
		\and Jae-Woo Kim \inst{4} \and Sang-Hyun Chun \inst{1}
		\and Ho-Il Kim \inst{3} \and Y.-J. Sohn \inst{1,2}}
\offprints{Jongwan Ko, \email{jwko@yonsei.ac.kr}}

%% ----------
%% Addresses
%% ----------
\institute{Department of Astronomy, Yonsei University, Seoul 120-749, Korea
		\and Yonsei University Observatory, Seoul 120-749, Korea
		\and Korea Astronomy and Space Science Institute, Daejeon 305-348, Korea
		\and CEOU, Department of Physics and Astronomy, 
			Seoul National University, Seoul 151-742, Korea}

%% --------------------------------
%% Dates of receipt and acceptance
%% --------------------------------
%\date{Received 2 November 1992 / Accepted 7 January 1993}

%% -----------------------------------------------
%% ABSTRACT - Aims, Methods, Results, Conclusions 
%% -----------------------------------------------
%% <example>
%%
%% \abstract {} {We look for characteristics typical of water-megamaser galaxies
%% in SO 103-G035, TXS 2226-184, and IC 1481.} {We obtained long-slit optical 
%% emission-line spectra.} {We present rotation curves, line ratios, electron 
%% densities, temperatures. IC 1481 reveals a spectrum suggestive of a vigorous 
%% starburst in the central kiloparsec 108 years ago.} {We do not find any hints 
%% for outflows nor special features which could give clues to the unknown 
%% megamaser excitation mechanism.} 

\abstract {} {We investigated the distribution of resolved asymptotic giant branch (AGB) 
stars over a much larger area than covered by previous near-infrared studies in the 
nearby dwarf elliptical galaxy NGC 205.} 
{Using data obtained with the WIRCam near-infrared imager of the CFHT, we selected the AGB stars 
in the $JHK_{s}$ color-magnitude diagrams, and separated the C stars from M-giant 
stars in the $JHK_{s}$ color-color diagram.} 
{We identified $1,550$ C stars in NGC 205 with a mean absolute magnitude of 
$\langle M_{K_s} \rangle$ = $-7.49 \pm 0.54$, 
and colors of $\langle (J-K_s)_{0} \rangle$ = $1.81 \pm 0.41$ and 
$\langle (H-K_s)_{0} \rangle$ = $0.76 \pm 0.24$.
The ratio of C stars to M-giant stars was estimated to be $0.15 \pm 0.01$ in NGC 205, 
and the local C/M ratios for the southern region are somewhat lower than those for 
the northern region. The ($J-K_{s}$) color distributions of AGB stars contain 
the main peak of the M-giant stars and the red tail of the C stars. 
A comparison of the theoretical isochrone models with the observed color distribution indicates 
that most of the bright M-giant stars in NGC 205 were formed at log($t_\mathrm{yr}$) $\sim 9.0$-$9.7$. 
The logarithmic slope of the $M_{K_s}$ luminosity function for M-giant stars was estimated to be 
$0.84 \pm 0.01$, which is comparable with dwarf elliptical galaxies NGC 147 and NGC 185. 
Furthermore, we found that the logarithmic slopes of the $M_{K_s}$ luminosity function 
for C and M-giant stars are different to places, implying a different star formation history within 
NGC 205.
The bolometric luminosity function for M-giant stars extends to $M_\mathrm{bol} = -6.0$ mag, 
and that for C stars spans $-5.6 < M_\mathrm{bol} < -3.0$. The bolometric luminosity function 
of C stars is unlikely to be a Gaussian distribution and the mean bolometric magnitude of C stars 
is estimated to be $M_\mathrm{bol} = -4.24 \pm 0.55$, which is consistent with our results for 
dwarf elliptical galaxies NGC 147 and NGC 185.} {}

%% ----------
%% Key words
%% ----------
%% <example>
%%
%%	\keywords{interstellar medium: jets and outflows --
%%  	interstellar medium: molecules -- stars: pre-main-sequence}} 


\maketitle 

%% ----------------------------------------------------- End manuscript header
%% ---------------------------------------------------------------------------
%% ---------------------------------------------------------------------------

%% ---------------------------------------------------------------------------
%% ---------------------------------------------------------------------------
%% 1. Introduction -----------------------------------------------------------

\section{Introduction} \label{sec:sec1}

Dwarf galaxies are the most numerous galactic systems in the Universe, 
and most are found in galaxy groups. 
Photometric studies of stars resolved in dwarf galaxies provide clues 
for stellar evolution and star formation history in the galaxies. 
Dwarf galaxies in the Local Group (LG) are therefore excellent laboratories 
in which to study galaxy evolution. 
Among the bright resolved stars in LG dwarf galaxies, asymptotic giant branch (AGB) stars 
allow us to study the late evolutionary stages of stars with low and intermediate masses
($\sim 0.8$-$8$ $M_{\sun}$) as well as the 
star formation history in a galaxy of the intermediate age of 
$1$-$10$ Gyr \citepads{1999IAUS..192...17G, 2003Ap&SS.284..579T}.

Stars in the AGB phase experience long-period pulsation variabilities,
large mass loss through stellar winds, and a change in atmospheric
chemical composition. During the thermal pulses (TP-AGB), material from
the nuclear region is mixed into the outer layers 
\citepads[dredge-up process;][]{1983ARA&A..21..271I}.
Owing to this dredge-up, AGB evolution can be classified according to atmospheric abundance,  
i.e., starting with oxygen-rich (C/O$<$1) M-giant stars that evolve into 
carbon-rich (C/O$>$1) C stars \citepads{2001A&A...367..557N}. These two types of AGB stars
can be used to estimate the metallicity of the parent galaxy because the ratio 
of C stars to M-giants (C/M) is anti-correlated with the metal abundance
\citepads{2005A&A...434..657B, 2005A&A...429..837C, 2009A&A...506.1137C}.
The AGB stars are among the brightest and coolest stars of the intermediate-age stellar population;
accordingly, they appear mostly in near-infrared color-magnitude diagrams (CMDs).
Furthermore, C stars are generally located in an extended red tail on near-infrared CMDs,
and can be easily and clearly distinguished from M-giant AGB stars
\citepads{1990AJ.....99..784H, 2000ApJ...542..804N, 2003A&A...402..133C, 
2005A&A...429..837C, 2003ApJ...597..289D, 2005AJ....130.2087D, 2005A&A...437...61K, 
2006A&A...454..717K, 2006A&A...445...69S, 2007A&A...466..501V}.
However, \citetads{2007A&A...474...35B} pointed out that near-infrared photometry only 
detects cool C stars and misses warm C stars, which can only be detected 
spectroscopically.

This paper is part of a series that studies the photometric properties of
AGB stars in LG dwarf galaxies from near-infrared photometry. 
In previous papers \citepads{2005A&A...437...61K, 2006A&A...454..717K, 2006A&A...445...69S}, 
we showed the near-infrared photometric properties of AGB stars in two dwarf elliptical
galaxies (NGC 185 and NGC 147) and a dwarf irregular galaxy (NGC 6822). 
We also demonstrated that C stars can be distinguished from AGB stars using near-infrared colors.
\citetads{2005A&A...437...61K} and \citetads{2006A&A...445...69S} provided 
near-infrared photometric properties of $73$ and $91$ C stars in NGC 185 and NGC 147,
respectively, and suggested that star formation in the galaxies has a wide range of
ages for intermediate-age stars. In this paper, we investigate
the near-infrared properties of resolved AGB populations in NGC 205,
a dwarf elliptical satellite of a large spiral galaxy (M31). 

Of the three dwarf elliptical companions of M31, 
NGC 205 has the shortest projected distance from the larger galaxy.
It is therefore expected that NGC 205 has experienced a much stronger interaction
with M31 than NGC 147 and NGC 185, and consequently has evolved differently. 
Indeed, many studies support this idea based on photometric and kinematics observations: 
the presence of young blue stars in the central region of NGC 205 \citepads[e.g.,][]{1999ApJ...515L..17C};
AGB stars in NGC 205 that show multiple episodes of star formation during the past gigayear 
\citepads{2003ApJ...597..289D}; the twisted surface isophotes and a subtle downward break 
in the surface brightness profile in NGC 205 \citepads{2002AJ....124..310C};
the presence of stellar tidal debris possibly associated with NGC 205 
\citepads{2003AJ....125.3037D, 2001Natur.412...49I, 2004MNRAS.351L..94M};
the dynamically distinct behavior of gas in NGC 205 
\citepads{1997ApJ...476..127Y, 1998ApJ...499..209W};
the estimation of the dust mass associated with the last burst of star formation in NGC 205 
\citepads{1991ApJ...374L..17F, 1998A&A...337L...1H, 2006ApJ...646..929M};
a steadily increasing velocity dispersion with radius \citepads{2002A&A...384..371S};
and a turnover in the major-axis velocity profile of NGC 205 at a radius of $4.5 \arcmin$ 
\citepads[$\sim 1$ kpc;][]{2006AJ....131..332G}.
To summarize, NGC 205 contains young stars, gas, and dust, whicj all imply recent star formation 
within a $1 \arcmin$ radius, beyond which the galaxy is free from gas and dust. This area also contains 
intermediate-age stars. NGC 205 furthermore shows the typical surface brightness and kinematics of a 
normal dwarf elliptical galaxy within an effective radius ($\sim 2.5 \arcmin$). 
However, beyond this radius, there is strong evidence of tidal distortion in the photometry 
and kinematics of NGC 205 \citepads{2002AJ....124..310C, 2006AJ....131..332G}.

Even though there is a strong interaction between M31 and NGC 205, the process that triggered 
star formation in the central region of NGC 205 is still being debated.
Some studies concluded that the latest star formation activities that occurred 
a few $10^{8}$ yr ago in the central region of NGC 205 might have been triggered 
by past interactions with M31 \citepads{2003ApJ...597..289D, 2005AJ....129.2217B},
while \citetads{2009A&A...502L...9M} recently argued against the tidally triggered star formation;
instead they suggested a constant star formation, at least over the past approximately $650$ Myr.
Moreover, \citetads{2008ApJ...683..722H} apparently interpreted the continuous star formation 
as being in line with the latest simulations of the NGC 205 orbit, 
where the authors found that the galaxy had its first interaction 
with M31, tangentially moving with a high velocity of $300$-$500$ km s$^{-1}$ 
from northwest to southeast.

Several studies have been devoted to investigating the stellar contents of 
resolved AGB and C stars toward NGC 205. \citetads{2003AJ....125.3037D} applied the four-band
photometric technique \citepads[proposed by][]{1986ApJ...305..634C} 
using $CN$ and $TiO$ narrow bands to identify C stars and found $532$ C stars 
in $42 \arcmin \times 28 \arcmin$ fields centered on NGC 205.
The estimated C/M ratio of NGC 205 in a $10 \arcmin$ ellipse is $0.09 \pm 0.01$ with $289$ C stars.
\citetads{2003ApJ...597..289D} has detected $320$ C stars from $JHK \arcmin$ images of 
$3.6 \arcmin \times 3.6 \arcmin$ field in the center of NGC 205. These authors found AGB stars
formed within the past $0.1$ Gyr and the previous episode of star formation
occurred a few tenths of a Gyr earlier, which is consistent with star formation
in NGC 205 being triggered by interactions with M31. 
Although \citetads{2005AJ....130.2087D} also used intermediate-age stars within an effective radius
and different color criteria to identify C stars, in this study we collected 
near-infrared photometric data for stars in a much wider area 
to study the properties of intermediate-age stars that reside in the tidally distorted region.  

In this paper, we investigate the bright AGB stellar populations of the
central $\sim 24 \arcmin \times 24 \arcmin$ of the nearby dwarf elliptical galaxy 
NGC 205 through near-infrared photometry. Section~\ref{sec:sec2} presents $JHK_{s}$ 
observations of NGC 205, the data reduction procedure, and photometric measurements of 
the resolved stars. The CMDs and color-color diagrams (CCDs) of resolved stars 
based on $JHK_{s}$ photometry are presented in Sect.~\ref{sec:sec3}. 
In Sect.~\ref{sec:sec4}, we discuss the near-infrared photometric properties of 
the identified C stars along with the distribution of the C/M ratio, 
the color distributions, and the luminosity functions of the AGB stars. 
A summary of the results is given in Sect.~\ref{sec:sec5}.

%% ---------------------------------------------------------------- End Chap 1 
%% ---------------------------------------------------------------------------
%% ---------------------------------------------------------------------------

%% star fig1  ----------------------------------------------------------------
\begin{figure}
	\resizebox{\hsize}{!}{\includegraphics{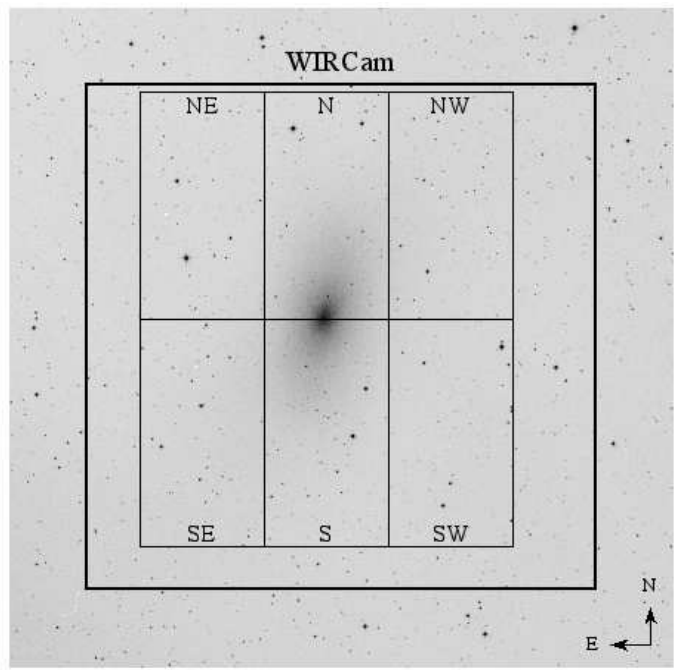}}
	\caption{Digitized Sky Survey image of NGC 205.
			The whole field-of-view is $32 \arcmin \times 32 \arcmin$ 
			with north up and east to the left. The WIRCam field covers a
			$\sim 24 \arcmin \times 24 \arcmin$ area in the 
			central area of NGC 205 (thick line). The six regions of 
			$\sim 6 \arcmin \times 11 \arcmin$ are divided to investigate 
			the spatial dependence of the photometric properties of 
			C stars in NGC 205 (thin line).}
	\label{fig:fig1}
\end{figure}
%% ------------------------------------------------------------------ end fig1

%% star tab1 -----------------------------------------------------------------
\begin{table}
	\caption{Observational log of July 19, 2007}            
	\label{tab:tab1}
	\centering
	\begin{tabular}{c c c c} 
	\hline\hline
 	Filter 	        & $\lambda_\mathrm{c}^\mathrm{eff}$ & Exp. time & FWHM	\\   
			& [ $\mu$m ]	& [ sec ]	& [ \arcsec ]	\\
	\hline
 	$J$		& $1.25$  & $4 \times 5 \times 30$	& $0.55$  \\
   	$H$     	& $1.63$  & $4 \times 5 \times 15$	& $0.52$  \\
   	$K_{s}$ 	& $2.15$  & $4 \times 5 \times 25$	& $0.53$  \\
	\hline
	\end{tabular}
\end{table}
%% ------------------------------------------------------------------ end tab1

%% ---------------------------------------------------------------------------
%% ---------------------------------------------------------------------------
%% 2. Observation, data reduction, photometry & AST test ---------------------

\section{Observation, data reduction, photometry, 
	and photometric measurements} \label{sec:sec2}

We obtained $JHK_{s}$ images via Queued Service Observing (QSO) with the WIRCam 
\citepads[a wide-field near-infrared detector;][]{2004SPIE.5492..978P}
at the $3.6$ m CFHT on July $19$, $2007$. The WIRCam contains four $2048 \times 2048$ 
HgCdTe arrays with a sampling of $0.3 \arcsec$ per pixel, so that each image covers 
a total field-of-view of $\sim 21 \arcmin \times 21 \arcmin$ of the sky. 
The total integration times on the central field of NGC 205 were $600$s ($J$), 
$300$s ($H$) and $500$s ($K_{s}$), which were obtained from individual exposures of 
$30$s ($J$), $15$s ($H$) and $25$s ($K_{s}$) with a five-point dithering pattern 
by $90 \arcsec$ offset to fill-in the $45 \arcsec$ gaps between arrays. 
At each pointing, the data were acquired using four half-pixel micro-dithering 
to improve the spatial resolution. Figure~\ref{fig:fig1} shows the observed fields, 
and a summary of the observation is presented in Table~\ref{tab:tab1}.

Each image was detrended (dark-subtraction, flat-fielding, etc.) using the 
`I`iwi\footnote{http://cfht.hawaii.edu/Instruments/Imaging/WIRCam/IiwiVersion1\\Doc.html}
preprocessing pipeline provided by the QSO team. After we subtracted the sky background level 
from each detrended image, we subtracted the thermal signatures using median 
combined flat-fielded images of blank sky regions. 
Then, we used TERAFIX\footnote{http://terapix.iap.fr/} software products 
(SExtractor, SCAMP, WeightWatcher, SWarp) to correct the distortion and chip-to-chip variation, 
and then co-added into one final mosaic image with the full exposure time of 
a $\sim 24 \arcmin \times 24 \arcmin$ field-of-view.
The seeings measured from the final images were about $0.52 \arcsec$-$0.55 \arcsec$ 
FWHM in $J$, $H$ and $K_{s}$.

We used the point-spread-function (PSF) fitting routines DAOPHOT II/ALLSTAR 
\citepads{1988AJ.....96..909S, 1987PASP...99..191S} to measure stellar brightness. 
We adopted a threshold value of the DAOPHOT sharp parameter 
(sharpness of 0.3-1.4) to weed out extended objects in the photometry, 
although this is complicated at the faint part.
The instrumental magnitudes of $JHK_{s}$ were photometrically calibrated 
using stars with $14 < K_{s} < 15$ in the 2MASS Point Source Catalog 
\citepads{2003tmc..book.....C} over the same area. 
A total of $145,246$ stars were detected in $JHK_{s}$ filters
on the $\sim 24 \arcmin \times 24 \arcmin$ field of the WIRCam observations.

\begin{figure}
	\includegraphics[width=9cm]{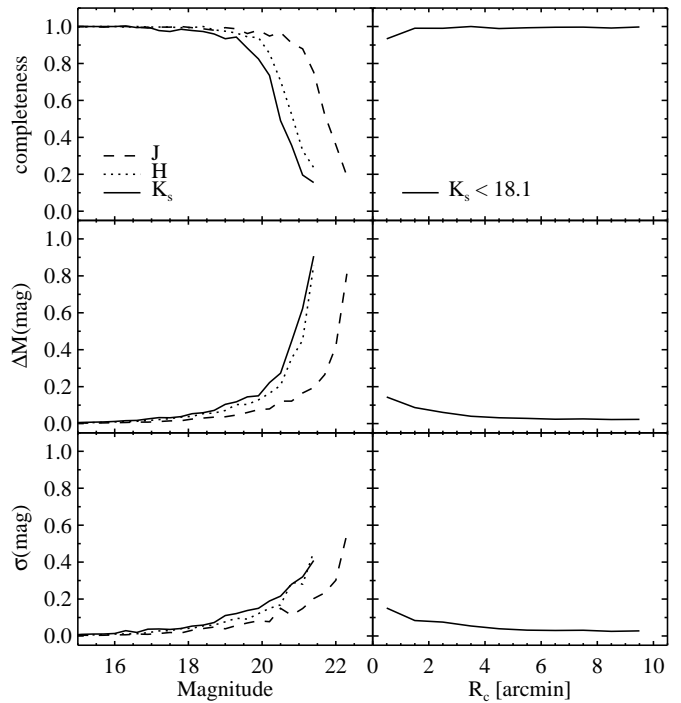}
	\caption{       $Left$: The results of artificial star experiments on the stacked 
			$JHK_{s}$ images (dashed, dotted, and solid lines) of a $\sim 24 \arcmin \times 24 \arcmin$ field of 
                        the WIRCam observation.			
			$Right$: The results of the experiments as a function of the projected 
                        radius ($R_{c}$) from the center of NGC 205. Only stars brighter than 
                        $K_{s}$=18.1 mag were considered because we assumed that AGB stars are brighter 
                        than $K_{s}$=18.1 mag in Sec. 3.2.
			The completeness is the number of recovered artificial stars divided
			by the total number of added stars.	$\Delta M$ (mag) is the mean
			difference between the actual input magnitudes and the measured
			magnitudes by DAOPHOTII/ALLSTARS, and $\sigma$ (mag) is the
			standard deviation of the $\Delta M$.
			}
	\label{fig:fig2}
\end{figure}

Completeness fractions and uncertainties in the photometric measurements
were estimated by performing artificial star experiments on the $JHK_{s}$ images. 
Only $400$ artificial stars were added to each experiment 
to take into account increasing uncertainties such as stellar density increases.
In the left panel of Fig.~\ref{fig:fig2}, a total of $10,000$ artificial stars 
from 25 experiments were added to the final mosaic $JHK_{s}$ images with a field-of-view of 
$24 \arcmin \times 24 \arcmin$.
Each horizontal panel in Fig.~\ref{fig:fig2} shows the completeness 
fraction defined as the recovery rate of the input artificial stars, 
the rms difference in magnitudes ($\Delta M$) between the measured and the input brightness, 
and the standard deviation of $\Delta M$ ($\sigma$).
When $J$, $H$ and $K_{s}$ are fainter than $20.9$, $20.0$ and $19.5$ mag, respectively,
the completeness is less than $\sim90 \%$.
To check the completeness difference as a function of stellar density, 
we also performed 36 new experiments by adding 400 artificial stars of $K_{s}$=15.5-18.5 mag
in each experiment with increasing projected radius from the center of NGC 205
($\alpha = 00^{\mathrm h} 40^{\mathrm m} 22.08^{\mathrm s}$, 
$\delta=+41 \degr 41\arcmin 07.10 \arcsec$). 
The right panel of Fig.~\ref{fig:fig2} shows the results of the experiments 
as a function of the projected radius ($R_{c}$) from the center of NGC 205. 
We only considered stars brighter than $K_{s} =18.1$ mag because we assumed 
that AGB stars are brighter than $K_{s}$=18.1 mag in Sec. 3.2. Although the completeness
slightly decreases in the central area ($R_{c} <1.5 \arcmin$), it is not less than $\sim90 \%$.
This suggests that incompleteness is not significant for our analysis.

%% ---------------------------------------------------------------- End Chap 2
%% ---------------------------------------------------------------------------
%% ---------------------------------------------------------------------------

%% star fig2  ----------------------------------------------------------------
%% ------------------------------------------------------------------ end fig2

%% star fig3  ----------------------------------------------------------------
\begin{figure*}
	\centering
	\includegraphics[width=17cm]{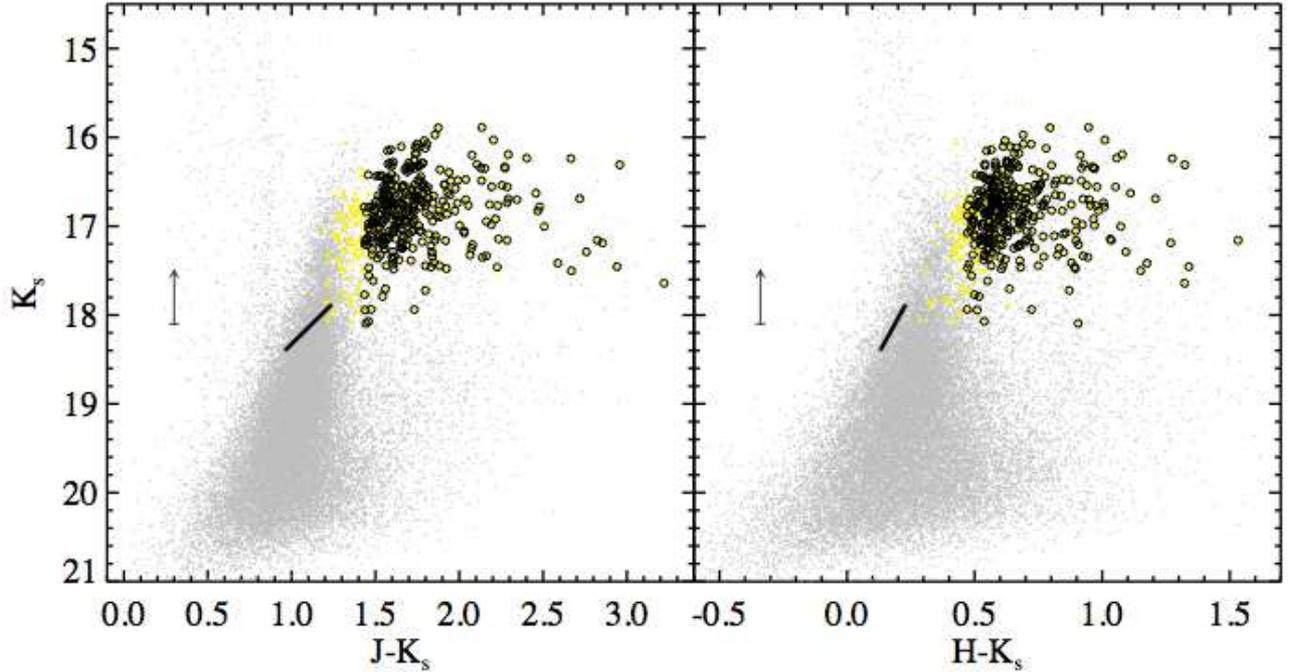}
	\caption{$(J-K_{s}, K_{s})$ and $(H-K_{s}, K_{s})$ CMDs for stars detected 
			in $JHK_{s}$ bands of the total WIRCam observed area 
			(only $20\%$ of stars randomly selected in the total sample are shown 
                        with gray dots).
			The isochrones are limited to the maximum mass-loss rate on the TP-AGB 
			with no-dust case.
			The estimated color and magnitude range of the TRGB 
			\citepads{2004MNRAS.354..815V} at the distance of NGC 205 also 
			are presented with thick slanted lines. Arrows indicate the estimated 
			TRGB magnitude in $K_{s}$ with $K_{s}=18.1$.
			The yellow plus symbol denotes C stars selected in $RICNTiO$ photometry 
			\citepads{2003AJ....125.3037D}, and the open circles represent 
			$309$ C stars cross-identified in $JHK_{s}$ photometry (in this work) 
			and $RICNTiO$ photometry \citepads{2003AJ....125.3037D}.}
	\label{fig:fig3}
\end{figure*}
%% ------------------------------------------------------------------ end fig3

%% ---------------------------------------------------------------------------
%% ---------------------------------------------------------------------------
%% 3. Photometric properties of AGB stars in NGC 205 -------------------------

\section{Photometric properties of AGB stars} \label{sec:sec3}

%% ---------------------------------------------------------------------------
%% 3.1 Adopted reddening, distance, and distance -----------------------------

\subsection{Adopted reddening, distance, and 
	metallicity range} \label{sec:sec3subsec1}

NGC 205 is the closest dwarf elliptical galaxy to M31 (located at $37 \arcmin$ from the center of M31) 
and behind M31 \citepads{2003AJ....125.1926G,2008ApJ...683..722H};  
the accurate color excess toward NGC 205 is not known yet because of M31's dust clouds. 
\citetads{1998ApJ...500..525S} estimated the foreground reddening (generated by the Galaxy) 
of $E(B-V)=0.083$ from the dust infrared emission feature. 
Adopting this value and applying the relative extinction ratios of 
\citetads{1998ApJ...500..525S}, we estimated the interstellar absorptions in $JHK_{s}$ 
passbands to be $A_{J} = 0.075$, $A_{H} = 0.048$, and $A_{K_{s}} = 0.030$. 
The calculated reddening values are then $E(J-K_{s}) = 0.045$ and $E(H-K_{s}) = 0.018$.

\citetads{2005MNRAS.356..979M} measured the distance modulus of NGC 205 as 
$(m-M)_{0} = 24.58$, based on the $I$-band tip of red-giant branch (TRGB) brightness. 
From investigating the interaction between NGC 205 and M31 with restricted 
$N$-body simulations, \citetads{2008ApJ...683..722H} estimated a distance modulus 
of $(m-M)_{0} = 24.50$. However, we adopted an unweighted mean distance modulus 
of NGC 205 as $(m-M)_0 = 24.54 \pm 0.09$ \citepads{2000glg..book.....V}.

\citetads{1984ApJ...278..575M} measured a metallicity of [Fe/H] $\gtrsim -0.9 \pm 0.2$
and a metallicity dispersion of $\pm 0.5$ dex, based on the color of the RGB stars in NGC 205. 
However, \citetads{1996ApJ...466..742J} estimated a much lower nuclear metallicity 
of [Fe/H] $\sim -1.4$ from the far-UV imaging data. 
\citetads{2005MNRAS.356..979M} also obtained the median metallicity of [Fe/H] = $-0.8$ 
(using theoretical evolutionary tracks with no $\alpha$-enhancement) for the RGB stars in NGC 205. 
Using $HST$ observations of the central region of NGC 205, \citetads{2005AJ....129.2217B} 
inferred the median metallicity of [Fe/H] $\gtrsim -1.06 \pm 0.04$ 
from ancient stars' color information, reaching [Fe/H] $\gtrsim -0.7$ for the RGB stars
that were the most metal-rich. The central region has also been observed spectroscopically. 
\citetads{1990A&A...228...23B} showed that the dominant population is young (a few $10^{8}$ years) 
with a maximum metallicity of [Fe/H] = $-0.5$, and much older stars with [Fe/H] $\gtrsim -1.0$. 
\citetads{2006MNRAS.372.1259S}, from a spectroscopic study of integrated light from stars
in the central regions of NGC 205, found a mean metallicity of [Fe/H] $\sim-0.5$ with 
a wide spread of metallicity $\sim 0.3$-$0.4$ dex. These studies indicate that 
NGC 205 has widespread metallicity in the range of [Fe/H] = $-1.4$ to $-0.5$.
Unless the age-metallicity relation in NGC 205 is peculiar, 
intermediate-age stars in NGC 205 should have metallicities that 
tend toward the higher end of the metallicity range found in the literature.

%% -------------------------------------------------------------- End Chap 3.1
%% ---------------------------------------------------------------------------

%% star fig4  ----------------------------------------------------------------
\begin{figure}
	\includegraphics{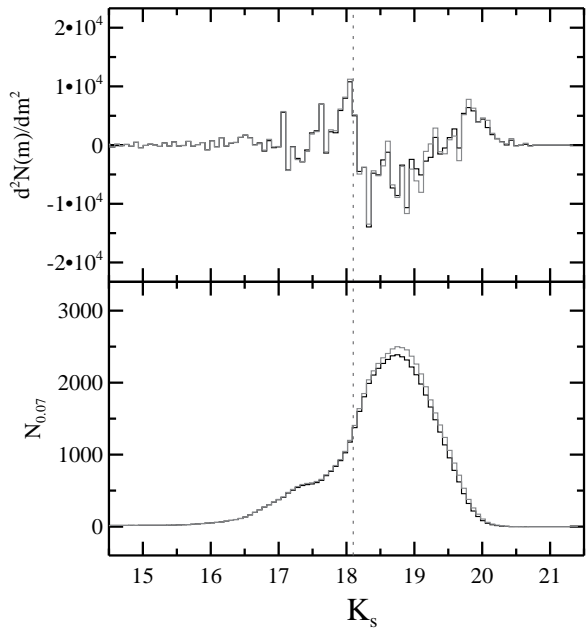}
	\caption{       {\it Lower}: The $K_{s}$ band luminosity function (LF) of 
			the resolved stars in NGC 205. Black line is the LF for the measured stars, 
			and the gray line is that for the completeness-corrected number of stars.
			{\it Upper}: The second derivatives of the observed LF and the completeness-corrected LF.
			Vertical dotted lines in each panel indicate the determined magnitudes of the TRGB 
			(i.e., $K_{s} = 18.1$ mag).
			}
	\label{fig:fig4}
\end{figure}
%% ------------------------------------------------------------------ end fig2

%% ---------------------------------------------------------------------------
%% 3.2 CMDs and selection of AGB stars ---------------------------------------
\subsection{Color-magnitude diagrams (CMDs)
	 and selection of AGB stars} \label{sec:sec3subsec2}

	Figure~\ref{fig:fig3} shows near-infrared ($J-K_{s}, K_{s}$) and ($H-K_{s}, K_{s}$) 
CMDs for resolved stars in the entire observed area of NGC 205. 
Using the empirical relations between TRGB brightness in near-infrared bands and 
metallicity for stars in a galaxy \citepads{2004MNRAS.354..815V},
we estimated the range of the magnitude and color of the TRGB for the given
metallicity range $[$Fe/H$]$ = $ -1.4 \sim -0.5$ of NGC 205. 
The thick lines in Fig.~\ref{fig:fig3} are the range of $K_{s}$ magnitudes and 
the colors of the TRGB at the adopted distance and metallicity range of NGC 205
($17.88 < K_{s} < 18.40$, $0.96 < (J-K_{s}) < 1.24$, and $0.13 < (H-K_{s}) < 0.23$).

To confirm the discontinuity in star counts near the expected location of the TRGB, 
we applied the procedure of \citetads{2000A&A...359..601C}. 
The TRGB discontinuity causes a peak in the second derivative of the observed LF, 
which is derived by using a Savitzky-Golay filter. The filter yields for bin number $i$,
$[d^{2}N/dm^{2}]_{i} = \Sigma^{J}_{j=-J}c_{j}N(m)_{i+j}$, where $N(m)$ is number of stars with 
$m$ magnitude and the $c_{j}$ are Savitzky-Golay coefficients for the chosen value of $J$ 
and the desired derivative order $L = 2$. The filter fits a polynomial on the order of $M$ to the data points 
$N(m)_{j}$ with $j = i−J, ..., i+J$, and then evaluates the $L^{\mathrm{th}}$ derivative of 
the polynomial at bin $i$ to estimate $d^{2}N/dm^{2}$.
By applying the Savitzky-Golay filter to the observed LF in $K_{s}$ band, 
we derived the second derivatives of LF per $0.07$ mag interval, and detected a peak 
of the second derivatives caused the TRGB discontinuity with $\sim$18.1  mag in $K_{s}$  
in Fig.~\ref{fig:fig4}.

From these approaches, we consider that AGB stars are located at brighter 
than $K_{s} = 18.1$ mag, and $11,928$ stars are accordingly selected as AGB stars.
The bright part for AGB stars has three different components in Fig.~\ref{fig:fig3}.
The bright blue sequence at $(J-K_{s}) < 0.97$ mostly corresponds to the Galactic foreground stars, 
and a vertical sequence ($0.97 < J-K_{s} < 1.43$) and a red plume ($J-K_{s} > 1.43$) 
to the M-giant and C stars \citepads{2003ApJ...597..289D}. 
The Galactic star count model of \citetads{1985ApJS...59...63R} predicted 
approximately $550$ bright blue foreground stars with $(B-V) < 1.3$ and $V<22$ 
in the $\sim 24 \arcmin \times 24 \arcmin$ region of the WIRCam observation toward NGC 205, 
which is roughly consistent with the count of the bright foreground stars in Fig.~\ref{fig:fig3}.
It should be noted that the faint part for AGB stars may be contaminated from 
unresolved background galaxies. According to \citetads[][see their Fig. 1]{2000MNRAS.311..707M}, there are
$\sim$4000 galaxies deg$^{2}$ when 16 $< K_{s} <$ 18.1 (magnitude range of AGB stars in Fig. 5).
This implies that $\sim$640 galaxies (16 $< K_{s} <$ 18.1) might be present in the WIRCam field.
In other words, at most $\sim$5\% of our AGB stars are probably intermediate-redshift galaxies.
We here assume that all types of background galaxies (16 $< K_{s} <$ 18.1)
will have $(J-K_{s})$ $\sim$ 1.0-2.0 from \citetads{1996AJ....112..839C}, applying the colors 
of galaxies and the k-corrections from \citetads{2001MNRAS.326..745M}.

\begin{figure}
	\includegraphics[scale=0.9]{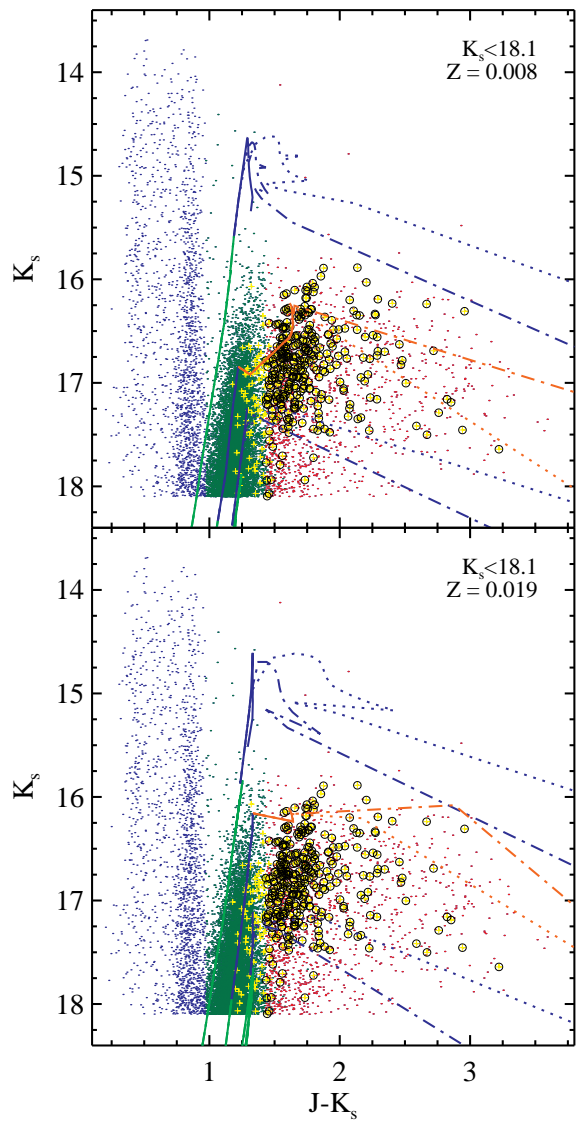}
	\caption{Theoretical isochrone of log$(t_\mathrm{yr})=8.2, 9.1$, and $10.0$
			with $Z=0.008$ ({\it upper}) and $Z=0.019$ ({\it lower}) 
			\citepads{2008A&A...482..883M} on the near-infrared ($J-K_{s},K_{s}$) 
			CMD of NGC 205. The different isochrone colors show different 
			evolutionary states. The green line corresponds to pre-TP-AGB stars, 
			and the blue and orange lines represent  M-giant stars and C stars, respectively. 
			The different dust obscuration models are represented by different line types;
			continuous lines for the no-dust, dot-dashed lines for \citetads{1998A&A...332..135B},
			and dashed lines for \citetads{2006A&A...448..181G} 
			(the $0.85$ AMC $+$ $0.15$ SiC mixture for C stars, 
			and $0.6$ silicate $+$ $0.4$ AlOx mixture for M-giant stars).
			The Galactic foreground stars with blue dots and the AGB star 
			(green: M-giants, and red: C stars) dots are divided at $(J-K_{s}) = 0.97$.
            The yellow plus and open circle symbols are the same as those in Fig.~\ref{fig:fig3}.}
	\label{fig:fig5}
\end{figure}

\begin{figure}
	\includegraphics{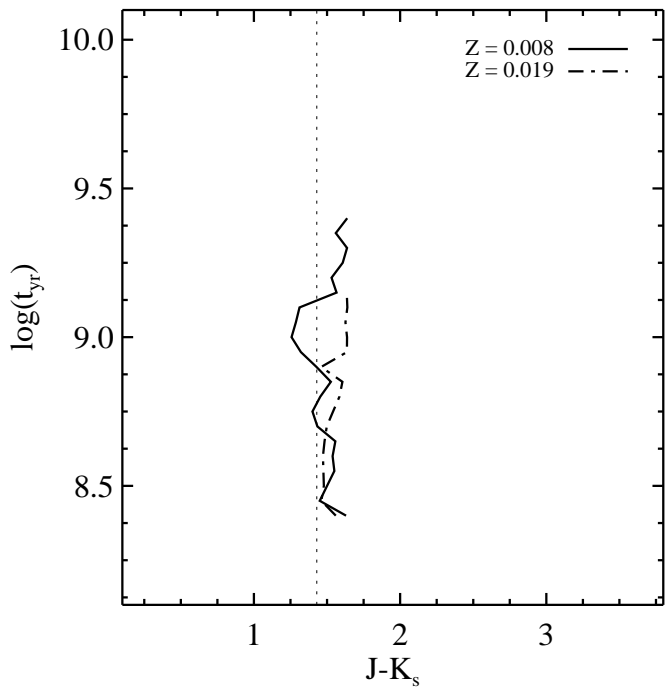}
	\caption{Function of age vs. the $(J-K_{s})$ color at the appearing C stars
			on the isochrones, i.e., C/O $> 1$, for the no-dust case \citepads{2008A&A...482..883M}. 
			The range of ages is log($t_\mathrm{yr}$) $= 8.40$-$9.40$ with $Z=0.008$ (solid line),
			and log($t_\mathrm{yr}$) $= 8.40$-$9.15$ with Z $= 0.019$ (dot-dashed line).
			The dotted line represents the adopted color cut of $(J-K_{s}) = 1.43$ 
			to separate C stars from M-giant stars.}
	\label{fig:fig6}
\end{figure}

To confirm the population classification, we compare the observed ($J-K_{s}, K_{s}$) CMD
shown in Fig.~\ref{fig:fig5} for stars brighter than $K_{s} = 18.1$ with 
the theoretical isochrones of AGB stars \citepads{2008A&A...482..883M}. 
We also plot the C stars identified in $RICNTiO$ photometry \citepads{2003AJ....125.3037D}.
Although C stars are broadly distributed over the CMD, most of the M-giants and C stars 
are clearly separated in the CMD \citepads[see Fig.~\ref{fig:fig7} in][]{2008A&A...482..883M}.
Figure~\ref{fig:fig6} shows ($J-K_{s}$) colors versus ages at the beginning of 
the C stars on the isochrones, i.e., C/O$>1$. The dotted line indicates 
the color cut of ($J-K_{s}$) $= 1.43$ to separate C stars from M-giants in the work.

%% -------------------------------------------------------------- End Chap 3.2
%% ---------------------------------------------------------------------------

%% star fig5  ----------------------------------------------------------------
%% ------------------------------------------------------------------ end fig5

%% star fig6  ----------------------------------------------------------------
%% ------------------------------------------------------------------ end fig6

%% ---------------------------------------------------------------------------
%% 3.3 CCDs and selection of C stars -----------------------------------------

\subsection{Color-color diagrams (CCDs) and 
	selection of C stars} \label{sec:sec3subsec3}
  
The narrowband photometric technique (i.e., $CN-TiO$ approach) for differentiating 
C stars from M-giants is based on different spectrum types of AGB stars, i.e., 
M-giants dominated by bands of the TiO molecule and C stars dominated by bands of 
the C$_{2}$ and CN molecules \citepads{2000AJ....119.2780A, 2000AJ....120.1801B, 
2001A&A...367..557N, 2003A&A...403...93N, 2002AJ....123..238D, 2002AJ....123..832L, 
2003AJ....125.1298B, 2003AJ....125.3037D, 2004A&A...418...33B, 2004A&A...417..479B, 
2004AJ....127.2711H, 2004A&A...427..613K}.
However, the extremely red C stars can be easily separated from the M-giants
because they have different locations on the near-infrared color-color diagrams
\citepads{1980ApJ...239..495F, 1985ApJ...290..477W, 2003A&A...403..225M, 
2005AJ....130.2087D, 2005A&A...437...61K, 2006A&A...454..717K, 2006A&A...445...69S}.

\begin{figure}
	\resizebox{\hsize}{!}{\includegraphics{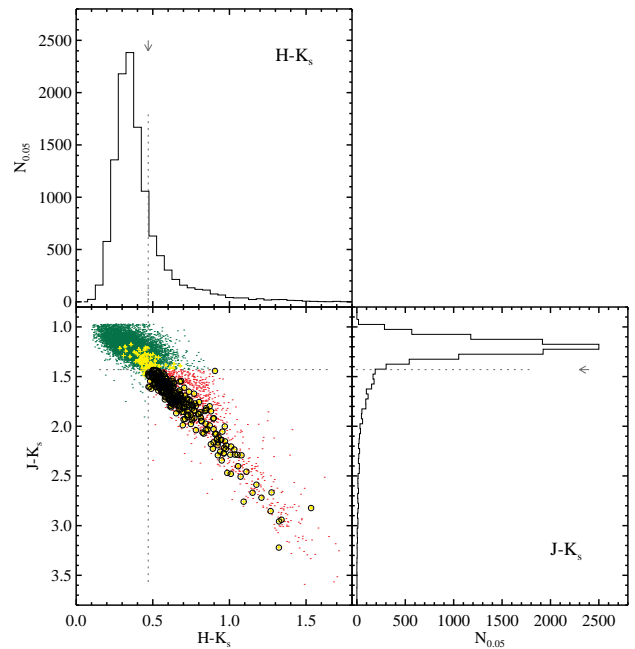}}
	\caption{$(H-K_s)$-$(J-K_s)$ color-color diagram and color histograms of 
			$(H-K_s)$ ({\it upper}) and $(J-K_s)$ ({\it right})
			for selected AGB stars. The dotted lines and arrows represent
			the color cuts of $(J-K_s)=1.43$ and $(H-K_s)=0.47$
			to separate C stars from M-giants.
			The green and red dots correspond to M-giants and C stars, respectively. 
			The yellow plus and open circle symbols are the same as those 
			in Fig.~\ref{fig:fig3} and Fig.~\ref{fig:fig5}.}
	\label{fig:fig7}
\end{figure}

\citetads{2003AJ....125.3037D} identified $532$ C stars in NGC 205 from their position 
on the $(CN-TiO)$ versus ($R-I$) CCD, using a CFH12K survey covering a field of view of 
$48 \arcmin \times28 \arcmin$ in NGC 205. 
By applying their criterion for selecting C stars 
(i.e., C stars: $(R-I) > 0.97, (CN-TiO) > 0.3$, M-giants: $(R-I) > 0.97, (CN-TiO) < 0.0$),
we found $394$ C stars and $5,054$ M-giant stars with $K_{s}$ magnitude brighter than $18.1$ 
in our WIRCam field.
We note, however, that it may possibly contain fainter stars than TRGB brightness.
\citetads{2005AJ....130.2087D} assigned the color limits of $(J-K_{s})_{0}>1.4$ and 
$(H-K_{s})_{0}>0.45$ to detach C stars from M-giants in nearby dwarf galaxies.
Adopting these color cuts to NGC 205, C stars were separated from M-giants at $(J-K_{s})>1.43$ 
and $(H-K_{s})>0.47$. Figure~\ref{fig:fig7} shows the $(H-K_{s})-(J-K_{s})$ CCD
and color histograms of selected AGB stars with the color cuts of C stars.
Most C stars selected in $RICNTiO$ photometry are also distributed in the color range 
of C stars in the $(H-K_{s})-(J-K_{s})$ CCD, and the color distribution shows 
the main peak of M-giants and the red tail of C stars. Finally, $1,550$ C stars were selected 
from the CCD among the observed $11,928$ AGB stars of NGC 205. Among the C stars selected in $JHK_{s}$, 
$309$ C stars matched $RICNTiO$ photometry \citepads{2003AJ....125.3037D}.
The cross-identified C stars are marked with open circles in
Figs.~\ref{fig:fig3},~\ref{fig:fig5}, and ~\ref{fig:fig7}. 
We can speculate that there is an obvious deficiency of Demers's C 
stars near the center of NGC 205 due to crowding, as Demers et al. (2003) also 
mentioned in their Figure 6. The seeing in $JHK_{s}$ images (0.52$\arcsec$-0.55$\arcsec$) 
was better than in $RICNTiO$ images (0.8$\arcsec$), and Demers et al. (2003) also mentioned that sky
conditions during the observations were nonphotometric. Therefore, more C stars 
could be discovered in $JHK_{s}$ images. Furthermore, Davidge (2005)
found 387 C stars near the center of NGC 205 ($3.7 \arcmin \times3.7 \arcmin$) using 
CFHTIR $JHK$ data, although they adopted a different faint limit for C stars than we did.
Therefore, the main reason for the difference in C stars compared to Demers and coworkers is that
they missed several C stars near the center of NGC 205.

%% -------------------------------------------------------------- End Chap 3.3
%% ---------------------------------------------------------------------------

%% ---------------------------------------------------------------- End Chap 3
%% ---------------------------------------------------------------------------
%% ---------------------------------------------------------------------------

%% star fig7  ----------------------------------------------------------------
%% ------------------------------------------------------------------ end fig7

%% ---------------------------------------------------------------------------
%% ---------------------------------------------------------------------------
%% ---------------------------------------------------------------------------
%% 4. Discussion
\section{Discussion} \label{sec:sec4}

%% ---------------------------------------------------------------------------
%% 4.1 C stars and the distribution of C/M ratio -----------------------------

\subsection{C stars and the distribution 
	of the C/M ratio} \label{sec:sec4subsec1}

The mean magnitude and colors of the $1,550$ C stars, 
which are classified in the $JHK_{s}$ bands,
are estimated to be $\langle K_{s} \rangle = 17.08 \pm 0.54$, 
$\langle (J-K_{s}) \rangle = 1.85 \pm 0.41$, 
and $\langle (H-K_{s}) \rangle = 0.78 \pm 0.24$. 
The corrected values are $\langle M_{K_{s}} \rangle = -7.49$, 
$\langle (J-K_{s})_{0} \rangle = 1.81$, and $\langle (H-K_{s})_{0} \rangle = 0.76$, 
with the distance modulus and reddening values in Sect.~\ref{sec:sec3subsec1}.
For the cross-identified $309$ C stars detected in common in the $RICNTiO$ bands 
by \citetads{2003AJ....125.3037D} and $JHK_{s}$ bands in this study, 
the mean absolute magnitude and colors are estimated to be 
$\langle M_{K_{s}} \rangle = -7.73 \pm 0.40$, $\langle (J-K_{s})_{0} \rangle = 1.72 \pm 0.30$, 
and $\langle (H-K_{s})_{0} \rangle = 0.67 \pm 0.18$. 
In our previous studies, the estimated mean absolute magnitude in the $K$ band of 
C stars in nearby dwarf galaxies, $\langle M_{K} \rangle$ is $-7.93 \pm 0.38$ 
for $73$ C stars in NGC 185 \citepads{2005A&A...437...61K}, 
$-7.56 \pm 0.47$ for $91$ C stars in NGC 147 \citepads{2006A&A...445...69S},
and $-7.60 \pm 0.50$ for $141$ C stars in NGC 6822 \citepads{2006A&A...454..717K}.

To investigate the spatial distribution of the C stars of NGC 205, 
we divided the observed WIRCam field of a $\sim 24 \arcmin \times 24 \arcmin$ area 
into six $\sim 6 \arcmin \times 11 \arcmin$ regions as shown in Fig.~\ref{fig:fig1}. 
Table~\ref{tab:tab2} summarizes the mean magnitude and colors for the C stars found in each region. 
The mean magnitude of the C stars is slightly brighter in the central part 
(i.e., `N' and `S' regions), with many more C stars than that of the C stars in the outer part.

%% star tab2  ----------------------------------------------------------------
\begin{table}
	\caption{The mean magnitude and colors of the selected $1,550$ C stars in NGC 205}
	\label{tab:tab2}
	\centering
	\begin{tabular}{c c c c c}       
	\hline\hline
	       &  C  &	$\langle K_{s} \rangle$ & $\langle (J-K_{s}) \rangle$ & $\langle (H-K_{s}) \rangle$ \\
	\hline
    NE                     &  $112$ &  $17.20 \pm  0.60$ &  $1.77 \pm  0.41$ &  $0.80 \pm  0.25$  \\
    N                      &  $397$ &  $16.90 \pm  0.44$ &  $1.87 \pm  0.41$ &  $0.75 \pm  0.24$  \\
	NW                     &  $114$ &  $17.15 \pm  0.44$ &  $1.87 \pm  0.42$ &  $0.81 \pm  0.22$  \\
    SE                     &  $184$ &  $17.22 \pm  0.53$ &  $1.89 \pm  0.49$ &  $0.81 \pm  0.28$  \\
    S                      &  $362$ &  $16.94 \pm  0.48$ &  $1.86 \pm  0.40$ &  $0.75 \pm  0.24$  \\
    SW                     &   $69$ &  $17.24 \pm  0.52$ &  $1.83 \pm  0.46$ &  $0.80 \pm  0.26$  \\
%    total\tablefootmark{1} & $1238$ &  $17.03 \pm  0.51$ &  $1.86 \pm  0.42$ &  $0.77 \pm  0.25$  \\
%    total\tablefootmark{2} & $1550$ &  $17.08 \pm  0.54$ &  $1.85 \pm  0.41$ &  $0.78 \pm  0.24$  \\
    total$^{1}$ 		   & $1238$ &  $17.03 \pm  0.51$ &  $1.86 \pm  0.42$ &  $0.77 \pm  0.25$  \\
    total$^{2}$ & $1550$ &  $17.08 \pm  0.54$ &  $1.85 \pm  0.41$ &  $0.78 \pm  0.24$  \\
	\hline                        
	\end{tabular}
%	\tablefoot{\\
%		\tablefoottext{1}{stars in $6$ selected regions} \\
%		\tablefoottext{2}{stars in the entire WIRCam observed area}}
	\begin{list}{}{}
		\item[$^{1}$] stars in six selected regions
		\item[$^{2}$] stars in the entire WIRCam observed area
	\end{list}
\end{table}
%% ------------------------------------------------------------------ end tab2

%% star tab3  ----------------------------------------------------------------
\begin{table}
	\caption{The number of C stars and M-giants, C/M ratio, and metallicity in NGC 205}
	\label{tab:tab3}
	\centering
	\begin{tabular}{c c c c c c}
	\hline\hline
%	    &  C  &  M  & C/M  & [Fe/H]\tablefootmark{a}  &	[Fe/H]\tablefootmark{b}  \\
	    &  C  &  M  & C/M  & [Fe/H]$^{a}$  &	[Fe/H]$^{b}$  \\
	\hline
	NE                       & $ 112$ & $  692$ &  $0.16 \pm  0.02$ & $-0.85 \pm  0.03$ & $-1.02 \pm  0.02$   \\
	N                        & $ 397$ & $ 2814$ &  $0.14 \pm  0.01$ & $-0.82 \pm  0.01$ & $-0.99 \pm  0.01$   \\
	NW                       & $ 114$ & $  718$ &  $0.16 \pm  0.02$ & $-0.85 \pm  0.03$ & $-1.01 \pm  0.02$   \\
	SE                       & $ 184$ & $ 1464$ &  $0.13 \pm  0.01$ & $-0.79 \pm  0.02$ & $-0.97 \pm  0.02$   \\
	S                        & $ 362$ & $ 2821$ &  $0.13 \pm  0.01$ & $-0.79 \pm  0.01$ & $-0.97 \pm  0.01$   \\
	SW                       & $  69$ & $  554$ &  $0.12 \pm  0.02$ & $-0.79 \pm  0.03$ & $-0.96 \pm  0.03$   \\
%	total\tablefootmark{1}   & $1238$ & $ 9063$ &  $0.14 \pm  0.01$ & $-0.81 \pm  0.01$ & $-0.98 \pm  0.01$   \\
%	total\tablefootmark{2}   & $1550$ & $10378$ &  $0.15 \pm  0.01$ & $-0.83 \pm  0.01$ & $-1.00 \pm  0.01$   \\
	total$^{1}$			     & $1238$ & $ 9063$ &  $0.14 \pm  0.01$ & $-0.81 \pm  0.01$ & $-0.98 \pm  0.01$   \\
	total$^{2}$			     & $1550$ & $10378$ &  $0.15 \pm  0.01$ & $-0.83 \pm  0.01$ & $-1.00 \pm  0.01$   \\
	\hline                        
	\end{tabular}
%	\tablefoot{ \\
%		\tablefoottext{a}{Calculated from the correlation of \citetads{2005A&A...434..657B}} \\
%		\tablefoottext{b}{Calculated from the correlation of \citetads{2009A&A...506.1137C}} \\
%		\tablefoottext{1}{stars in $6$ selected regions} \\
%		\tablefoottext{2}{stars in the entire WIRCam observed area}}
	\begin{list}{}{}
		\item[$^{a}$] Calculated from the correlation of \citetads{2005A&A...434..657B}
		\item[$^{b}$] Calculated from the correlation of \citetads{2009A&A...506.1137C}
		\item[$^{1}$] stars in six selected regions
		\item[$^{2}$] stars in the entire WIRCam observed area
	\end{list}
\end{table}
%% ---------------------------------------------------------- end tab3

We estimate the ratio of the C stars to M-giants (C/M) of the AGB population
to be $0.15 \pm 0.01$ at the WIRCam observed field of NGC 205. As seen in Table~\ref{tab:tab3}, 
we did not find any spatial difference in the local C/M ratio of NGC 205.
Instead, the mean C/M ratio in the southern part seems to be slightly lower than in the northern part.
This implies that the star formation history for AGB stars in NGC 205 is likely to be
different for the northern and southern parts.
Alternatively, the difference in C/M ratios between the northern and 
southern parts of NGC 205 might be caused by contamination from the disk 
of M31, which lies immediately to the south of NGC 205. The M31 disk is 
more metal-rich than NGC 205 and consequently will have a lower C/M ratio. 
Presumably, a significant fraction of the stars in the southern half 
of the NGC 205 field belong to the M31 disk.
\citetads{2003AJ....125.3037D} found $289$ C stars and evaluated the C/M ratio 
in the central $10 \arcmin$ ellipse of NGC 205 as $0.09 \pm 0.01$, 
using the $CN-TiO$ technique to identify C stars.
To examine the difference in the C/M ratios, we applied the color criterion of 
\citetads{2003AJ....125.3037D} (C stars: $(R-I)_{0} > 0.9$ and $(CN-TiO) > 0.3$;
M-giants: $(R-I)_{0} > 0.9$ and $(CN-TiO) < 0.0$) to our C and M-giant stars 
identified in $JHK_s$ photometry and found $394$ C stars and $5,054$ M-giants. 
The calculated C/M ratio in the observed WIRCam field of NGC 205 is $0.08 \pm 0.01$.
Therefore, the difference in the C/M ratio is largely due to the method for defining 
the M-giants and C stars of the AGB population.

The C/M ratio of the LG galaxies correlates with metallicity, for instance, a lower C/M ratio 
corresponds to a higher metallicity \citepads{1999IAUS..191..535G, 2002A&A...387..507M}.
\citetads{2005A&A...434..657B} obtained C/M ratios for the LG galaxies, 
and found the correlation [Fe/H]$^{a}$ = $-1.32(\pm 0.07) -0.59(\pm 0.09) \times$log(C/M). 
Updating the metallicities for LG galaxies, the correlation is calibrated as 
[Fe/H]$^{b}$ = $-1.39(\pm 0.06) -0.47(\pm 0.10) \times$log(C/M) \citepads{2009A&A...506.1137C}.
Using these correlations, we calculated [Fe/H] = $-0.79 \sim -1.02$ 
in the WIRCam observed area of NGC 205. Table~\ref{tab:tab3} summarizes 
the number of C stars and M-giants, C/M ratios, and metallicities of NGC 205. 
\citetads{2007A&A...474...35B} argue that applying C/M (obtained from $RICNTiO$) vs. [Fe/H] calibration 
to C/M (obtained from $JHK_{s}$) vs. [Fe/H] calibration is not justified. 
Therefore, a more uniform data set for the LG galaxies are needed to confirm 
the correlation in the near-infrared wavelengths.

In previous studies, the estimated C/M ratios of the two elliptical satellites of M31 
are $0.11 \pm 0.01$ for NGC 185 \citepads{2005A&A...437...61K} and 
$0.16 \pm 0.02$ for NGC 147 \citepads{2006A&A...445...69S}.
 And the derived metallicities [Fe/H] are $-0.76 \pm 0.03$ and $-0.86 \pm 0.03$ 
from \citetads{2005A&A...434..657B}, and $-0.95 \pm 0.03$ and $-1.02 \pm 0.02$ 
from \citetads{2009A&A...506.1137C}, for NGC 185 and NGC 147, respectively.
This indicates that the metallicity of NGC 205 is more similar to NGC 147 than to NGC 185.
However, the color criteria for selecting C stars from M-giants are different for each galaxy. 
Accordingly, we re-estimated the C stars of NGC 147 and NGC 185, using the color criteria for NGC 205
(i.e., $(J-K)_{0} > 1.4$ and $(H-K)_{0} > 0.45$). As a result,
the C/M ratios are $0.18 \pm 0.02$ in NGC 185 and $0.17 \pm 0.02$ in NGC 147. 
The derived metallicities [Fe/H] are then $-0.88 \pm 0.03$ and $-0.86 \pm 0.03$ 
from \citetads{2005A&A...434..657B}, and $-1.04 \pm 0.02$ and $-1.03 \pm 0.02$ 
from \citetads{2009A&A...506.1137C}, for NGC 185 and NGC 147, respectively.
This result suggests that the metallicities of the three dwarf elliptical satellites of M31 are not 
very different, although the observed areas of two galaxies are much smaller than NGC 205.
The relative metallicities of these three galaxies can also be compared from the 
luminosity-metallicity ($M_{B}$ vs. [Fe/H]) relation obtained by using oxygen abundances 
of HII regions for dwarf irregular galaxies \citepads{2006ApJ...637..269V}. 
The metallicities of NGC 147, 185, and 205 are estimated to be 
$-1.12$, $-1.14$, and $-0.94$, respectively. We here assumed that 12+log(O/H)$_{\odot}$=8.66 
\citepads{2003ASPC..304..275A} and used the relation [Fe/H]=[O/H]$-0.37$ \citepads{1998ARA&A..36..435M}. 
It should be noted that at a fixed luminosity, dwarf ellipticals have a higher metallicity than dwarf irregulars. 
Therefore, the derived metallicities from the relationship can be slightly 
underestimated \citepads[e.g.,][]{2007MNRAS.375..715G}.

%% -------------------------------------------------------------- End Chap 4.1
%% ---------------------------------------------------------------------------

%% star fig8  ----------------------------------------------------------------
\begin{figure*}
	\centering
	\includegraphics[width=17cm]{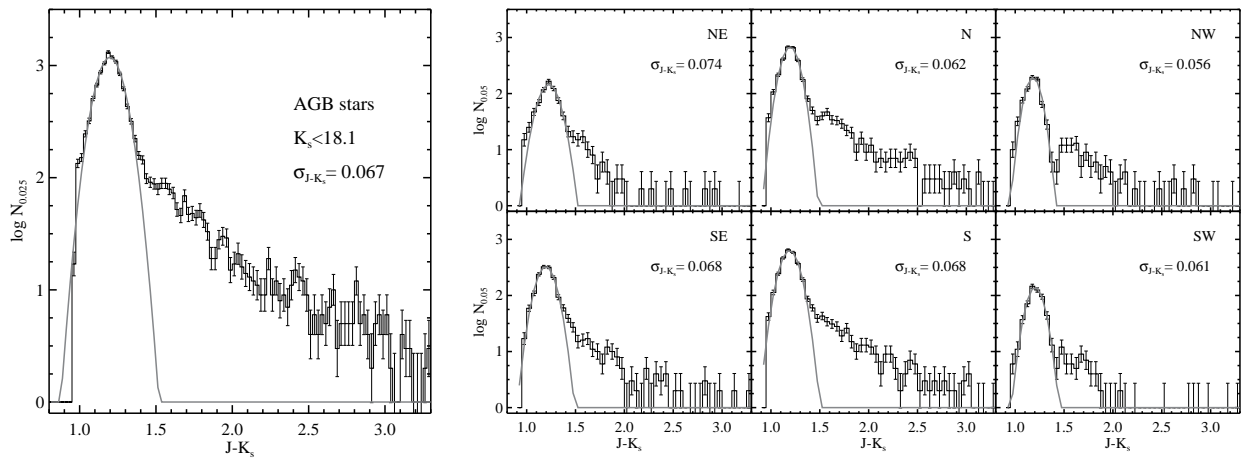}
	\caption{Histogram distributions of $(J-K_{s})$ colors		
			for the resolved $11,928$ AGB stars of the WIRCam fields ($left$)
			and the AGB stars in each region ($right$).
			The standard deviation ($\sigma_{J-K_{s}}$) is the $1 \sigma$ deviation 
                        estimated by the Gaussian fitting of the histogram.}
	\label{fig:fig8}
\end{figure*}
%% ------------------------------------------------------------------ end fig8 

%% ---------------------------------------------------------------------------
%%% 4.2 Color distributions

\subsection{Color distributions} \label{sec:sec4subsec2}

Figure~\ref{fig:fig8} shows the histogram distributions of the 
$(J-K_{s})$ color for the selected $11,928$ AGB stars ({\it left}), 
and for AGB stars belonging to each region ({\it right}) in NGC 205.
All histograms of $(J-K_{s})$ have the main peak of M-giant stars and the red tail of C stars.
The shapes of the color distributions for AGB stars in each region do not show 
significantly different features from the color distribution for all selected AGB stars. 
However, the standard deviation ($\sigma_{J-K_{s}}$) of the eastern part (i.e., `NE' and `SE') is 
larger than that of the western part (i.e. `NW' and `SW'), indicating a much wider range of 
stellar ages in the eastern area of NGC 205 because the color distribution of AGB stars 
can be used to infer the intermediate-age stellar populations in a galaxy 
\citepads{2005AJ....130.2087D, 2005A&A...437...61K, 2006A&A...454..717K, 2006A&A...445...69S}.
However, this difference is possibly due to contamination of
the color distributions from some foreground stars and/or stars belonging to M31.

In our previous studies \citepads{2005A&A...437...61K, 2006A&A...454..717K, 2006A&A...445...69S},
we used the color distributions of NGC 185, NGC 147, and NGC 6822 
to infer the range of ages for intermediate-aged populations in each galaxy:
two possible epochs in NGC 185 with log($t_\mathrm{yr}) \sim 9.0$-$9.4$ 
and $7.8$-$8.5$ \citepads{2005A&A...437...61K},
a range of epochs in NGC 147 with approximately $8.2 \la$ log($t_\mathrm{yr}$) $\la 8.6$ 
having a peak of log($t_\mathrm{yr}$) $= 8.4$ \citepads{2006A&A...445...69S},
a range of epochs in NGC 6822 with approximately $8.0 \la$ log($t_\mathrm{yr}$) $\la 10.0$ 
having a peak of log($t_\mathrm{yr}$) $= 9.0$ \citepads{2006A&A...454..717K}.
Similarly, \citetads{2005AJ....130.2087D} also inferred the range of ages for M-giant stars 
in NGC 147, NGC 185, and NGC 205 by comparing the ($J-K$) color in the color histograms of M-giant stars 
with $Z=0.008$ with the \citetads{2002A&A...391..195G} isochrones for $M_{K_{s}}=-7.4$ 
and ages of log($t_\mathrm{yr}$) $= 8.1, 9.0$, and $9.8$.
Davidge restricted the brightness range of M-giant stars with $-7.6 < M_{K} < -7.2$ 
to avoid sample incompleteness and to span the full range of ages of the stars. 

\begin{figure}
	\resizebox{\hsize}{!}{\includegraphics{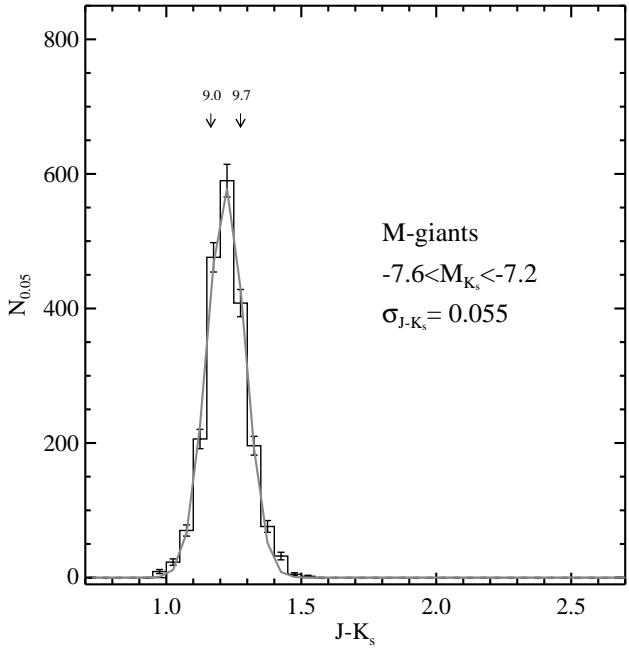}}
	\caption{Histogram distribution of ($J-K_{s}$) color for $1,630$ 
			M-giant stars with $-7.6 < M_{K_{s}} < -7.2$ per $0.05$ mag interval. 
			Arrows indicate the predicted ($J-K_{s}$) colors of the isochrones 
			for M-giant stars \citep{2008A&A...482..883M} at $M_{K_{s}} = -7.4$ 
			with $Z=0.008$ and ages of log($t_\mathrm{yr}$) $= 9.0$ and $9.7$.}
	\label{fig:fig9}
\end{figure}

To estimate the range of ages for M-giant stars in NGC 205, we also applied the same technique.
Figure~\ref{fig:fig9} shows the histogram distribution of the $(J-K_{s})$
for $1,630$ M-giant stars with $-7.6 < M_{K_{s}} < -7.2$ at an interval of $0.05$ mag.
Arrows are the ($J-K_{s}$) colors of the theoretical isochrones of M-giant stars 
at $M_{K_{s}} = -7.4$, with $Z = 0.008$ and no-dust case, 
for ages of log($t_\mathrm{yr}$) $= 9.0$ and $9.7$ with the adopted distance and reddening of NGC 205.
The spread of the color distribution of the M-giant stars indicates that 
NGC 205 contains M-giant stars spanning a wide range of star formation epochs 
(i.e., log$(t_\mathrm{yr}) \sim 9.0$-$9.7$), having a peak of log($t_\mathrm{yr}$) $= 9.3$.
Our WIRCam field is approximately $32$ times larger than that of \citetads{2005AJ....130.2087D}, and 
we used the new isochrones of \citetads{2008A&A...482..883M}.

%% -------------------------------------------------------------- End Chap 4.2
%% ---------------------------------------------------------------------------

%% star tab4  ------------------------------------------------------------------------
\begin{table*}
	\caption{The mean absolute and bolometric mangitude of C stars, 
			and the logarithmic slope of $M_{K_s}$ LF for C and M-giant stars in NGC 205
			}
	\label{tab:tab4}
	\centering
	\begin{tabular}{c c c c c}
	\hline\hline
	    & $\langle M_{K_{s}} \rangle_{\rm{C\ stars}}$ & $\langle M_\mathrm{bol} \rangle_{\rm{C\ stars}}$	&  slope for C stars & slope for M-giants  \\
		&							  &										& (-8.5 $ < M_{K_s} <$ -7.7) & (-8.5 $ < M_{K_s} <$ -7.1) \\
	\hline
	NE          & $-7.37 \pm 0.60$  & $-4.17 \pm 0.59$ & $0.64 \pm 0.13$ & $0.53 \pm  0.02$    \\
	N           & $-7.67 \pm 0.44$  & $-4.41 \pm 0.46$ & $1.03 \pm 0.02$ & $0.93 \pm  0.01$    \\
	NW          & $-7.42 \pm 0.44$  & $-4.17 \pm 0.46$ & $0.97 \pm 0.16$ & $1.45 \pm  0.03$    \\
	SE          & $-7.35 \pm 0.53$  & $-4.08 \pm 0.58$ & $0.59 \pm 0.11$ & $0.73 \pm  0.01$    \\
	S           & $-7.63 \pm 0.48$  & $-4.38 \pm 0.49$ & $0.68 \pm 0.03$ & $0.85 \pm  0.01$    \\
	SW          & $-7.33 \pm 0.52$  & $-4.09 \pm 0.52$ & $0.93 \pm 0.22$ & $0.73 \pm  0.03$    \\
	total$^{1}$ & $-7.54 \pm 0.51$  & $-4.29 \pm 0.52$ & $0.88 \pm 0.01$ & $0.88 \pm  0.01$    \\
	total$^{2}$ & $-7.49 \pm 0.54$  & $-4.24 \pm 0.55$ & $0.89 \pm 0.01$ & $0.84 \pm  0.01$    \\
	\hline                        
	\end{tabular}
	\begin{list}{}{}
		\item[$^{a}$] Calculated from the correlation of \citetads{2005A&A...434..657B}
		\item[$^{b}$] Calculated from the correlation of \citetads{2009A&A...506.1137C}
		\item[$^{1}$] stars in six selected regions
		\item[$^{2}$] stars in the entire WIRCam observed area
	\end{list}
\end{table*}
%% ---------------------------------------------------------- end tab4

%% ---------------------------------------------------------------------------
%% 4.3 Luminosity functions --------------------------------------------------

\subsection{Luminosity functions} \label{sec:sec4subsec3}

Luminosity functions (LFs) for the resolved AGB stars in a galaxy are 
a useful tool for comparing the evolution models of AGB stars with observations.
Moreover, bolometric LF and the mean bolometric magnitude of C stars in a galaxy
depend upon the metallicity and star formation history of a galaxy
\citepads{1999IAUS..191..535G, 2003A&A...403...93N}.
Near-infrared observations have an advantage of deriving the LFs of AGB stars
in a galaxy because some red bright stars detected on near-infrared images are not
always visible in optical images.

\begin{figure}
%	\centering
	\resizebox{\hsize}{!}{\includegraphics{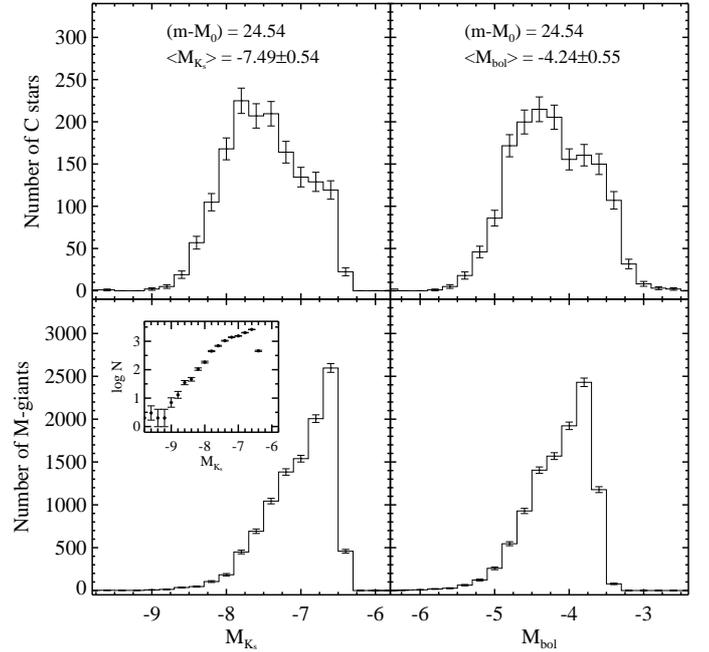}}
	\caption{{\it Left}: the completeness-corrected LFs in the $M_{K_{s}}$ 
			for the M-giants and C stars. The inner small panel in the lower
			panel indicates the logarithmic LF of M-giant of NGC 205.
			The LFs with solid lines show the M-giants and C stars detected in
			the WIRCam field of NGC 205.	
			{\it Right}: the completeness-corrected bolometric LFs
			for the M-giants and C stars.}
	\label{fig:fig10}
\end{figure}

The two diagrams on the left of Fig.~\ref{fig:fig10} show the completeness-corrected LFs 
in the $M_{K_{s}}$ for the M-giants and C stars detected in the entire observed area of NGC 205.
The completeness-corrected logarithmic LF in the $M_{K_{s}}$ of M-giant stars 
is also shown in the inner small panel in the lower-left panel of Fig.~\ref{fig:fig10}.
The $M_{K_{s}}$ LF of C stars is unlikely to be a Gaussian distribution, 
which is consistent with our previous results in NGC 147 and NGC 185. 
We estimated the logarithmic slope of  the $M_{K_{s}}$ LF for the M-giant stars in NGC 205 
to be $0.84 \pm 0.01$, by performing a least-squares fit to the brightness interval of 
$-8.5 < M_{K_{s}} < -7.1$. Compared to the slopes with our previous studies for nearby dwarf galaxies, 
this value is comparable with dwarf elliptical galaxies, i.e., $0.83 \pm 0.02$ 
for NGC 185 \citepads{2005A&A...437...61K} and, $0.79 \pm 0.02$ for NGC 147 
\citepads{2006A&A...445...69S}, but is lower than a dwarf irregular galaxy, i.e, 
$1.12 \pm 0.03$ for NGC 6822 \citepads{2006A&A...454..717K}.
This result implies that the star formation history for M-giant stars among dwarf elliptical galaxies 
is likely similar but different from that in the dwarf irregular galaxy. 
We will check whether the slope in the $K$ band LF of M-giants depends 
on the morphology of dwarf galaxies in a forthcoming paper.

Figure 11 shows the $M_{K_s}$ and $M_\mathrm{bol}$ LFs for the C stars in 
each region of NGC 205. We estimated the mean magnitude in the $K_{s}$ band and the logarithmic 
slope of the $M_{K_s}$ LF for the C stars in each region of NGC 205 (Table~\ref{tab:tab4}). 
For comparison, we also estimated the logarithmic slope of the $M_{K_s}$ LF for the M-giant stars in 
each region, and found that `NE' (0.53) and `NW' (1.45) differ from the value for total 
M-giants (0.84). It also seems that the slopes of the $M_{K_s}$ LF for the C stars in the northern 
part are slightly steeper than in the southern part. With the result of the higher C/M ratio in 
the northern part, this suggests that the star formation history is different to places within NGC 205.

Bolometric corrections in the $K$ band, $BC_{K}$, were estimated separately for the 
M-giants and C stars in NGC 205 from the empirical relations between $BC_{K}$ 
and $(J-K)$ for Galactic and LMC AGB stars. For the M-giants, we applied the relations 
for the Galactic and LMC M-giants given by \citetads{1984PASP...96..247B}, 
while for the C stars we applied that for the Galactic C stars given by \citetads{1996AJ....112.2607C}.
The right panels of Fig.~\ref{fig:fig10} show the completeness-corrected bolometric LFs of
the M-giants and C stars in NGC 205. The bolometric LF for the M-giant stars in NGC 205
extends to $M_\mathrm{bol} = -6.0$ mag, and that of the C stars spans $-5.6 < M_\mathrm{bol} < -3.0$. 
The mean bolometric LF of the C stars in NGC 205 is estimated to be $M_\mathrm{bol} = -4.24 \pm 0.55$.
This is the equivalent of those of C stars in nearby dwarf galaxies, 
i.e., $M_\mathrm{bol} = -4.50 \pm 0.42$ in NGC 185 \citepads{2005A&A...437...61K}, 
$M_\mathrm{bol} = -4.32 \pm 0.49$ in NGC 147 \citepads{2006A&A...445...69S}, 
and $M_\mathrm{bol} = -4.36 \pm 0.54$ in NGC 6822 \citepads{2006A&A...454..717K}.

%% -------------------------------------------------------------- End Chap 4.3
%% ---------------------------------------------------------------------------

%% ---------------------------------------------------------------- End Chap 4
%% ---------------------------------------------------------------------------
%% ---------------------------------------------------------------------------

%% star fig9 -----------------------------------------------------------------
%% ------------------------------------------------------------------ end fig9

%% star fig10 -----------------------------------------------------------------
%% ------------------------------------------------------------------ end fig10

%% ---------------------------------------------------------------------------
%% 5. Summary ----------------------------------------------------------------

\section{Summary} \label{sec:sec5}

The $JHK_{s}$ images were used to investigate the AGB population 
in the nearby dwarf elliptical galaxy NGC 205. 
Our main results are summarized as follows.

\begin{enumerate}
	\item The CMDs of NGC 205 based on the $JHK_{s}$ photometry
		show various populations of MS, RGB, AGB, and foreground stars.
		The AGB populations in the CMDs are dominated by M-giant stars and red C stars.	

	\item A total of $1,550$ C stars were selected from AGB stars in the
		near-infrared CCD. The mean magnitude and colors of
		the $1,550$ C stars in $\sim 24 \arcmin \times 24 \arcmin$ field of NGC 205
		are estimated to be $\langle K_{s} \rangle = 17.08 \pm 0.54$, 
		$\langle (J-K_{s}) \rangle = 1.85 \pm 0.41$
		and $\langle (H-K_{s}) \rangle = 0.78 \pm 0.24$. 
		This is equivalent to $\langle M_{K_{s}} \rangle = -7.49$,
		$\langle (J-K_{s})_{0} \rangle = 1.81$,
		and $\langle (H-K_{s})_{0} \rangle = 0.76$ at the distance of NGC 205.

	\item The C/M ratio of the entire observed WIRCam field is estimated to be
		$0.15 \pm 0.01$. The slightly lower C/M ratio (i.e., higher metallicity) in the 
		southern part compared to the northern part could be the result of different star 
		formation history in NGC 205. 

	\item The ($J-K_{s}$) color histogram shows the main peak of M-giant stars
		and the red tail of C stars. A spread of the color distribution of the M-giant stars
		indicates a range of star formation epochs log($t_\mathrm{yr}$) $\sim 9.0$-$9.7$ in NGC 205.

	\item The logarithmic slope of the $M_{K_{s}}$ LF for the M-giant stars with 
		$-8.5 < M_{K_{s}} < -7.1$ is estimated to be $0.84 \pm 0.01$. 
		This value is similar to dwarf elliptical galaxies NGC 147 and NGC 185, 
		but lower than the irregular galaxy NGC 6822.

	\item The logarithmic slopes of the $M_{K_s}$ LF for the C and M-giant stars
                are slightly different to places, implying a different 
                star formation history within NGC 205.

	\item The bolometric LF for M-giants in NGC 205 extends to $M_\mathrm{bol} = -6.0$
		mag, and that for C stars spans $-5.6 < M_\mathrm{bol} < -3.0$.	
		The bolometric LF of C stars is probably not a Gaussian distribution, 
		and the mean bolometric magnitude of C stars is estimated to be
		$M_\mathrm{bol} = -4.24 \pm 0.55$, which is comparable with our previous results for 
		the dwarf elliptical galaxies NGC 147 and NGC 185. 

\end{enumerate}

%% ---------------------------------------------------------------- End Chap 5
%% ---------------------------------------------------------------------------
%% ---------------------------------------------------------------------------

%% star fig11  ---------------------------------------------------------------
\begin{figure*}
	\centering
	\includegraphics[width=17cm]{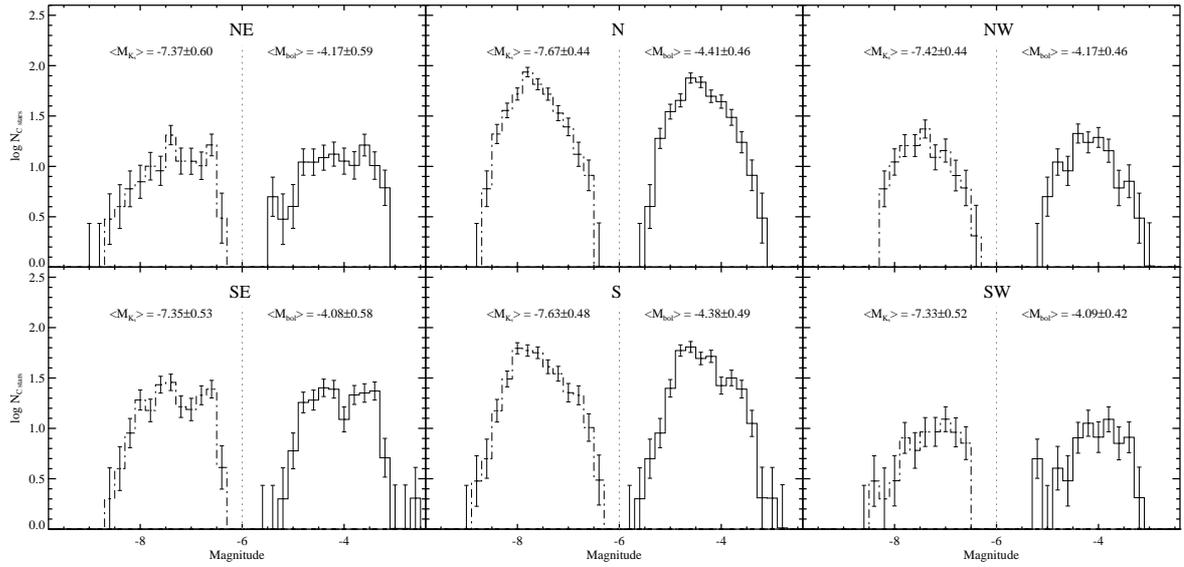}
	\caption{Completeness-corrected LFs in the $M_{K_s}$ (dot-dashed)
	and the $M_\mathrm{bol}$ (solid) for the C stars in each region.
	}
	\label{fig:fig11}
\end{figure*}
%% ----------------------------------------------------------------- end fig11
%% ---------------------------------------------------------------------------

%% ---------------------------------------------------------------------------
%% ---------------------------------------------------------------------------
%% Acknowledgements ----------------------------------------------------------

\begin{acknowledgements}

This research was supported by the Basic Science Research Program through the National
Research Foundation of Korea (NRF) funded by the Ministry of Education, Science and
Technology (2010-0021867). This work is partially supported by the KASI-Yonsei Joint
Research Program (2010-2011) for the Frontiers of Astronomy and Space Science funded 
by the Korea Astronomy and Space Science Institute.
M.Y. Jung is grateful for the support provided by the EU FP7 participation support
program (2008-2011) of the National Research Foundation of Korea.

\end{acknowledgements}

%% ------------------------------------------------------ End Acknowledgements
%% ---------------------------------------------------------------------------
%% ---------------------------------------------------------------------------

%% ---------------------------------------------------------------------------
%% ---------------------------------------------------------------------------
%% Reference -----------------------------------------------------------------
\bibliographystyle{aa} % style aa.bst
\bibliography{ngc205_aasty1} % your references Yourfile.bib

%%Query Results from the ADS Database
%%Retrieved 76 abstracts, starting with number 1.  Total number selected: 76.

\begin{thebibliography}{76}
\expandafter\ifx\csname natexlab\endcsname\relax\def\natexlab#1{#1}\fi


\bibitem[Albert et al.(2000)]{2000AJ....119.2780A} Albert, L., Demers, S., 
\& Kunkel, W.~E.\ 2000, \aj, 119, 2780 


\bibitem[Asplund(2003)]{2003ASPC..304..275A} Asplund, M.\ 2003, 
Astronomical Society of the Pacific Conference Series, 304, 275 


\bibitem[Battinelli et 
al.(2007)]{2007A&A...474...35B} Battinelli, P., Demers, S., \& Mannucci, F.\ 2007, \aap, 474, 35 


\bibitem[Battinelli 
\& Demers(2005)]{2005A&A...434..657B} Battinelli, P., \& Demers, S.\ 2005, \aap, 434, 657 


\bibitem[Battinelli 
\& Demers(2004a)]{2004A&A...418...33B} Battinelli, P., \& Demers, S.\ 2004a, \aap, 418, 33 


\bibitem[Battinelli 
\& Demers(2004b)]{2004A&A...417..479B} Battinelli, P., \& Demers, S.\ 2004b, \aap, 417, 479 


\bibitem[Battinelli et al.(2003)]{2003AJ....125.1298B} Battinelli, P., 
Demers, S., \& Letarte, B.\ 2003, \aj, 125, 1298 


\bibitem[Battinelli 
\& Demers(2000)]{2000AJ....120.1801B} Battinelli, P., \& Demers, S.\ 2000, \aj, 120, 1801 


\bibitem[Bessell 
\& Wood(1984)]{1984PASP...96..247B} Bessell, M.~S., \& Wood, P.~R.\ 1984, \pasp, 96, 247 


\bibitem[Bica et 
al.(1990)]{1990A&A...228...23B} Bica, E., Alloin, D., \& Schmidt, A.~A.\ 1990, \aap, 228, 23 


\bibitem[Bressan et 
al.(1998)]{1998A&A...332..135B} Bressan, A., Granato, G.~L., \& Silva, L.\ 1998, \aap, 332, 135 


\bibitem[Butler 
\& Mart{\'{\i}}nez-Delgado(2005)]{2005AJ....129.2217B} Butler, D.~J., \& Mart{\'{\i}}nez-Delgado, D.\ 2005, \aj, 129, 2217 


\bibitem[Cappellari et al.(1999)]{1999ApJ...515L..17C} Cappellari, M., 
Bertola, F., Burstein, D., et al.\ 1999, \apjl, 515, L17 


\bibitem[Choi et al.(2002)]{2002AJ....124..310C} Choi, P.~I., Guhathakurta, 
P., \& Johnston, K.~V.\ 2002, \aj, 124, 310 


\bibitem[Cioni(2009)]{2009A&A...506.1137C} Cioni, M.-R.~L.\ 2009, \aap, 506, 1137 


\bibitem[Cioni 
\& Habing(2005)]{2005A&A...429..837C} Cioni, M.-R.~L., \& Habing, H.~J.\ 2005, \aap, 429, 837 


\bibitem[Cioni 
\& Habing(2003)]{2003A&A...402..133C} Cioni, M.-R.~L., \& Habing, H.~J.\ 2003, \aap, 402, 133 


\bibitem[Cioni et 
al.(2000)]{2000A&A...359..601C} Cioni, M.-R.~L., van der Marel, R.~P., Loup, C., \& Habing, H.~J.\ 2000, \aap, 359, 601 


\bibitem[Cook et al.(1986)]{1986ApJ...305..634C} Cook, K.~H., Aaronson, M., 
\& Norris, J.\ 1986, \apj, 305, 634 


\bibitem[Costa 
\& Frogel(1996)]{1996AJ....112.2607C} Costa, E., \& Frogel, J.~A.\ 1996, \aj, 112, 2607 


\bibitem[Cowie et al.(1996)]{1996AJ....112..839C} Cowie, L.~L., Songaila, 
A., Hu, E.~M., \& Cohen, J.~G.\ 1996, \aj, 112, 839 


\bibitem[Cutri et al.(2003)]{2003tmc..book.....C} Cutri, R.~M., Skrutskie, 
M.~F., van Dyk, S., et al.\ 2003, ''The IRSA 2MASS All-Sky Point Source 
Catalog, NASA/IPAC Infrared Science 
%Archive.~<A href=''http://irsa.ipac.caltech.edu/applications/Gator/''>http://irsa.ipac.caltech.edu/applications/Gator/</A>'',  
Archive. http://irsa.ipac.caltech.edu/applications/Gator/''  


\bibitem[Davidge(2005)]{2005AJ....130.2087D} Davidge, T.~J.\ 2005, \aj, 
130, 2087 


\bibitem[Davidge(2003)]{2003ApJ...597..289D} Davidge, T.~J.\ 2003, \apj, 
597, 289 


\bibitem[Demers et al.(2003)]{2003AJ....125.3037D} Demers, S., Battinelli, 
P., \& Letarte, B.\ 2003, \aj, 125, 3037 


\bibitem[Demers 
\& Battinelli(2002)]{2002AJ....123..238D} Demers, S., \& Battinelli, P.\ 2002, \aj, 123, 238 


\bibitem[Fich 
\& Hodge(1991)]{1991ApJ...374L..17F} Fich, M., \& Hodge, P.\ 1991, \apjl, 374, L17 


\bibitem[Frogel et al.(1980)]{1980ApJ...239..495F} Frogel, J.~A., Persson, 
S.~E., \& Cohen, J.~G.\ 1980, \apj, 239, 495 


\bibitem[Geha et al.(2006)]{2006AJ....131..332G} Geha, M., Guhathakurta, 
P., Rich, R.~M., \& Cooper, M.~C.\ 2006, \aj, 131, 332 


\bibitem[Girardi et 
al.(2002)]{2002A&A...391..195G} Girardi, L., Bertelli, G., Bressan, A., et al.\ 2002, \aap, 391, 195 


\bibitem[Gon{\c c}alves et al.(2007)]{2007MNRAS.375..715G} Gon{\c c}alves, 
D.~R., Magrini, L., Leisy, P., \& Corradi, R.~L.~M.\ 2007, \mnras, 375, 715 


\bibitem[Grebel et al.(2003)]{2003AJ....125.1926G} Grebel, E.~K., 
Gallagher, J.~S., III, \& Harbeck, D.\ 2003, \aj, 125, 1926 


\bibitem[Grebel(1999)]{1999IAUS..192...17G} Grebel, E.~K.\ 1999, The 
Stellar Content of Local Group Galaxies, 192, 17 


\bibitem[Groenewegen(2006)]{2006A&A...448..181G} Groenewegen, M.~A.~T.\ 2006, \aap, 448, 181 


\bibitem[Groenewegen(1999)]{1999IAUS..191..535G} Groenewegen, M.~A.~T.\ 
1999, Asymptotic Giant Branch Stars, 191, 535 


\bibitem[Grebel et al.(2003)]{2003AJ....125.1926G} Grebel, E.~K., 
Gallagher, J.~S., III, \& Harbeck, D.\ 2003, \aj, 125, 1926 


\bibitem[Haas(1998)]{1998A&A...337L...1H} Haas, M.\ 1998, \aap, 337, L1 


\bibitem[Harbeck et al.(2004)]{2004AJ....127.2711H} Harbeck, D., Gallagher, 
J.~S., III, \& Grebel, E.~K.\ 2004, \aj, 127, 2711 


\bibitem[Howley et al.(2008)]{2008ApJ...683..722H} Howley, K.~M., Geha, M., 
Guhathakurta, P., et al.\ 2008, \apj, 683, 722 


\bibitem[Hughes 
\& Wood(1990)]{1990AJ.....99..784H} Hughes, S.~M.~G., \& Wood, P.~R.\ 1990, \aj, 99, 784 


\bibitem[Ibata et al.(2001)]{2001Natur.412...49I} Ibata, R., Irwin, M., 
Lewis, G., Ferguson, A.~M.~N., \& Tanvir, N.\ 2001, \nat, 412, 49 


\bibitem[Iben 
\& Renzini(1983)]{1983ARA&A..21..271I} Iben, I., Jr., \& Renzini, A.\ 1983, \araa, 21, 271 


\bibitem[Jones et al.(1996)]{1996ApJ...466..742J} Jones, D.~H., Mould, 
J.~R., Watson, A.~M., et al.\ 1996, \apj, 466, 742 


\bibitem[Kang et 
al.(2006)]{2006A&A...454..717K} Kang, A., Sohn, Y.-J., Kim, H.-I., et al.\ 2006, \aap, 454, 717 


\bibitem[Kang et 
al.(2005)]{2005A&A...437...61K} Kang, A., Sohn, Y.-J., Rhee, J., et al.\ 2005, \aap, 437, 61 


\bibitem[Kerschbaum et 
al.(2004)]{2004A&A...427..613K} Kerschbaum, F., Nowotny, W., Olofsson, H., \& Schwarz, H.~E.\ 2004, \aap, 427, 613 


\bibitem[Letarte et al.(2002)]{2002AJ....123..832L} Letarte, B., Demers, 
S., Battinelli, P., \& Kunkel, W.~E.\ 2002, \aj, 123, 832 


\bibitem[Mannucci et al.(2001)]{2001MNRAS.326..745M} Mannucci, F., Basile, 
F., Poggianti, B.~M., et al.\ 2001, \mnras, 326, 745 


\bibitem[Marigo et 
al.(2008)]{2008A&A...482..883M} Marigo, P., Girardi, L., Bressan, A., et al.\ 2008, \aap, 482, 883 


\bibitem[Marigo et 
al.(2003)]{2003A&A...403..225M} Marigo, P., Girardi, L., \& Chiosi, C.\ 2003, \aap, 403, 225 


\bibitem[Marigo(2002)]{2002A&A...387..507M} Marigo, P.\ 2002, \aap, 387, 507 


\bibitem[Marleau et al.(2006)]{2006ApJ...646..929M} Marleau, F.~R., 
Noriega-Crespo, A., Misselt, K.~A., et al.\ 2006, \apj, 646, 929 


\bibitem[Mateo(1998)]{1998ARA&A..36..435M} Mateo, M.~L.\ 1998, \araa, 36, 435 


\bibitem[McConnachie et al.(2005)]{2005MNRAS.356..979M} McConnachie, A.~W., 
Irwin, M.~J., Ferguson, A.~M.~N., et al.\ 2005, \mnras, 356, 979 


\bibitem[McConnachie et al.(2004)]{2004MNRAS.351L..94M} McConnachie, A.~W., 
Irwin, M.~J., Lewis, G.~F., et al.\ 2004, \mnras, 351, L94 


\bibitem[McCracken et al.(2000)]{2000MNRAS.311..707M} McCracken, H.~J., 
Metcalfe, N., Shanks, T., et al.\ 2000, \mnras, 311, 707 


\bibitem[Monaco et 
al.(2009)]{2009A&A...502L...9M} Monaco, L., Saviane, I., Perina, S., et al.\ 2009, \aap, 502, L9 


\bibitem[Mould et al.(1984)]{1984ApJ...278..575M} Mould, J., Kristian, J., 
\& Da Costa, G.~S.\ 1984, \apj, 278, 575 


\bibitem[Nikolaev 
\& Weinberg(2000)]{2000ApJ...542..804N} Nikolaev, S., \& Weinberg, M.~D.\ 2000, \apj, 542, 804 


\bibitem[Nowotny et 
al.(2003)]{2003A&A...403...93N} Nowotny, W., Kerschbaum, F., Olofsson, H., \& Schwarz, H.~E.\ 2003, \aap, 403, 93 


\bibitem[Nowotny et 
al.(2001)]{2001A&A...367..557N} Nowotny, W., Kerschbaum, F., Schwarz, H.~E., \& Olofsson, H.\ 2001, \aap, 367, 557 


\bibitem[Puget et al.(2004)]{2004SPIE.5492..978P} Puget, P., Stadler, E., 
Doyon, R., et al.\ 2004, \procspie, 5492, 978 


\bibitem[Ratnatunga 
\& Bahcall(1985)]{1985ApJS...59...63R} Ratnatunga, K.~U., \& Bahcall, J.~N.\ 1985, \apjs, 59, 63 


\bibitem[Schlegel et al.(1998)]{1998ApJ...500..525S} Schlegel, D.~J., 
Finkbeiner, D.~P., \& Davis, M.\ 1998, \apj, 500, 525 


\bibitem[Sharina et al.(2006)]{2006MNRAS.372.1259S} Sharina, M.~E., 
Afanasiev, V.~L., \& Puzia, T.~H.\ 2006, \mnras, 372, 1259 


\bibitem[Simien 
\& Prugniel(2002)]{2002A&A...384..371S} Simien, F., \& Prugniel, P.\ 2002, \aap, 384, 371 


\bibitem[Sohn et 
al.(2006)]{2006A&A...445...69S} Sohn, Y.-J., Kang, A., Rhee, J., et al.\ 2006, \aap, 445, 69 


\bibitem[Stetson 
\& Harris(1988)]{1988AJ.....96..909S} Stetson, P.~B., \& Harris, W.~E.\ 1988, \aj, 96, 909 


\bibitem[Stetson(1987)]{1987PASP...99..191S} Stetson, P.~B.\ 1987, \pasp, 
99, 191 


\bibitem[Tolstoy(2003)]{2003Ap&SS.284..579T} Tolstoy, E.\ 2003, \apss, 284, 579 


\bibitem[Valcheva et 
al.(2007)]{2007A&A...466..501V} Valcheva, A.~T., Ivanov, V.~D., Ovcharov, E.~P., \& Nedialkov, P.~L.\ 2007, \aap, 466, 501 


\bibitem[Valenti et al.(2004)]{2004MNRAS.354..815V} Valenti, E., Ferraro, 
F.~R., \& Origlia, L.\ 2004, \mnras, 354, 815 


\bibitem[van den Bergh(2000)]{2000glg..book.....V} van den Bergh, S.\ 2000, 
The galaxies of the Local Group, by Sidney Van den Bergh.~Published by 
Cambridge, UK: Cambridge University Press, 2000 Cambridge Astrophysics 
Series Series, vol no: 35, ISBN: 0521651816.,  


\bibitem[van Zee et al.(2006)]{2006ApJ...637..269V} van Zee, L., Skillman, 
E.~D., \& Haynes, M.~P.\ 2006, \apj, 637, 269 


\bibitem[Welch et al.(1998)]{1998ApJ...499..209W} Welch, G.~A., Sage, 
L.~J., \& Mitchell, G.~F.\ 1998, \apj, 499, 209 


\bibitem[Wood et al.(1985)]{1985ApJ...290..477W} Wood, P.~R., Bessell, 
M.~S., \& Paltoglou, G.\ 1985, \apj, 290, 477 


\bibitem[Young 
\& Lo(1997)]{1997ApJ...476..127Y} Young, L.~M., \& Lo, K.~Y.\ 1997, \apj, 476, 127 


\end{thebibliography}
%% ---------------------------------------------------------------------------
%% ------------------------------------------------------------- End Reference
%% ---------------------------------------------------------------------------

\end{document}